\documentclass[aps,pra,reprint,superscriptaddress,showkeys,amsmath,amssymb,longbibliography]{revtex4-1}

\usepackage[english]{babel}
\usepackage{graphicx}% Include figure files
\usepackage{dcolumn}% Align table columns on decimal point
\usepackage{bm}% bold math
\usepackage{cellspace}%
\usepackage{array}%
\usepackage{
	xcolor,
	xspace,
	siunitx,
	braket,
    ragged2e,
    boldline
}

\usepackage{hyperref}% add hypertext capabilities
\usepackage[xspace]{ellipsis}

\newcommand{\ie}{i.\,e.\ }
\newcommand{\eg}{e.\,g.\ }
\newcommand{\cf}{cf.\ }
\newcommand{\Rb}{$^{87}$Rb\xspace}
\makeatletter    % Macro for balanced page break
\newcommand*{\balancecolsandclearpage}{%
  \close@column@grid
  \clearpage
  \twocolumngrid
}
\makeatother

\DeclareSIUnit\gauss{G}
\DeclareSIUnit\photons{photons}
\DeclareSIUnit\atoms{atoms}
\begin{document}
\title{
 Remote optimization of an ultra-cold atoms experiment by experts and citizen scientists
}

\author{Robert Heck}
\author{Oana Vuculescu}
\author{Jens Jakob S\o{}rensen}
\affiliation{ScienceAtHome, Department of Physics and Astronomy, Aarhus University, Aarhus, Denmark}
\author{Jonathan Zoller}
\affiliation{IQST, Ulm University, Ulm, Germany}
\author{Morten G. Andreasen}
\affiliation{ScienceAtHome, Department of Physics and Astronomy, Aarhus University, Aarhus, Denmark}
\author{Mark G. Bason}
\affiliation{Department of Physics and Astronomy, University of Sussex, Falmer, Brighton, United Kingdom}
\author{Poul Ejlertsen} 
\author{Ott\'{o} El\'{i}asson}
\author{Pinja Haikka}
\author{Jens S. Laustsen}
\author{L\ae{}rke L. Nielsen}
\affiliation{ScienceAtHome, Department of Physics and Astronomy, Aarhus University, Aarhus, Denmark}
\author{Andrew Mao}
\author{Romain M\"u{}ller}
\author{Mario Napolitano}
\author{Mads K. Pedersen}
\author{Aske R. Thorsen}
\author{Carsten Bergenholtz}
\affiliation{ScienceAtHome, Department of Physics and Astronomy, Aarhus University, Aarhus, Denmark}
\author{Tommaso Calarco}
\affiliation{IQST, Ulm University, Ulm, Germany}
\author{Simone Montangero}
\affiliation{IQST, Ulm University, Ulm, Germany}
\affiliation{Theoretische Physik, Universit\"a{}t des Saarlandes, Saarbr\"u{}cken, Germany}
\affiliation{Dipartimento di Fisica e Astronomia, Universit\`{a} degli Studi di Padova, Italy}
\author{Jacob F. Sherson}\email[Electronic address: ]{sherson@phys.au.dk}
\affiliation{ScienceAtHome, Department of Physics and Astronomy, Aarhus University, Aarhus, Denmark}

\date{\today}% It is always \today, today,

\begin{abstract}
We introduce a novel remote interface to control and optimize the experimental production of Bose-Einstein condensates (BECs) and find improved solutions using two distinct implementations. First, a team of theoreticians employed a Remote version of their dCRAB optimization algorithm (RedCRAB), and second a gamified interface allowed 600 citizen scientists from around the world to participate in real-time optimization. Quantitative studies of player search behavior demonstrated that they collectively engage in a combination of local and global search. This form of \textit{adaptive search} prevents premature convergence by the explorative behavior of low-performing players while high-performing players locally refine their solutions. In addition, many successful citizen science games have relied on a problem representation that directly engaged the visual or experiential intuition of the players. Here we demonstrate that citizen scientists can also be successful in an entirely abstract problem visualization. This gives encouragement that a much wider range of challenges could potentially be open to gamification in the future. 
\end{abstract}


\maketitle

In modern scientific research, high-tech applications, such as quantum computation~\cite{ladd_quantum_2010}, require exquisite levels of control while taking into account increasingly complex environmental interactions~\cite{devitt_quantum_2013}. This necessitates continual development of optimization methodologies. The fitness landscape~\cite{Wright1932}, spanned by the possible controls and their corresponding solution quality, forms a unifying mathematical framework for search problems in both natural~\cite{Kryazhimskiy2009,Rabitz2004,Malan2014,Rabitz2014,Sorensen2016,palittapongarnpim2016learning} and social science~\cite{Levinthal1997,acerbi2016social}.
Generally, search in the landscape can be approached with local or global optimization methods. Local solvers are analogous to greedy hill climbers and global methods attempt to investigate the entire landscape using stochastic steps.  
Achieving the proper balance between these is often referred to as the exploration/exploitation trade-off in both machine learning (ML)~\cite{dueck_new_1993} and social sciences \cite{march1991exploration}.

Much effort in computer science is therefore focused on developing algorithms that exploit the topology of the landscape to adapt search strategies and make better-informed jumps ~\cite{Malan2014,mersmann2011exploratory}. ML algorithms have achieved success across numerous domains. However, among researchers pursuing truly domain-general artificial intelligence, there is a growing call to rely on insights from human behavior and psychology \cite{lake2016building,Marcus2018}. 
Thus, emphasis is currently shifting towards the development of human-machine hybrid intelligence \cite{Kamar2016,baltz_achievement_2017}.
     
At the same time quantum technology is starting to step out of university labs into the corporate world. For the realization of real-world applications, not only must hardware be improved but also proper interfaces and software need to be developed. Examples of such interfaces are the \textit{IBM quantum experience}~\cite{ibm2016} and \textit{Quantum in the Cloud}~\cite{quantumcloud}, which give access to their quantum computing facilities and have ushered in an era in which theoreticians can experimentally test and develop their error correction models and new algorithms directly~\cite{Hebenstreit2017}. 
The optimal development of such interfaces, allowing the smooth  transformation of human intuition or experience-based insights (heuristics) into algorithmic advances, necessitates understanding and explicitly formulating the search strategies introduced by the human expert.
     
The emerging field of citizen science provides a promising way to investigate and harness the unique problem-solving abilities humans possess~\cite{bonney2014}. In recent years,  the creativity and intuition of non-experts using gamified interfaces have enabled scientific contributions across different fields such as quantum physics~\cite{Sorensen2016}, astrophysics~\cite{Lintott2008} and computational biology~\cite{Cooper2010,Lee2014c,Kim2014}. Here, citizen scientists often seem to jump across very rugged landscapes and solve non-convex optimization problems efficiently using search methodologies that are difficult to quantify and encode in a computer algorithm.

\begin{figure}[t]
     	\centering
     	\includegraphics[width=1\linewidth]{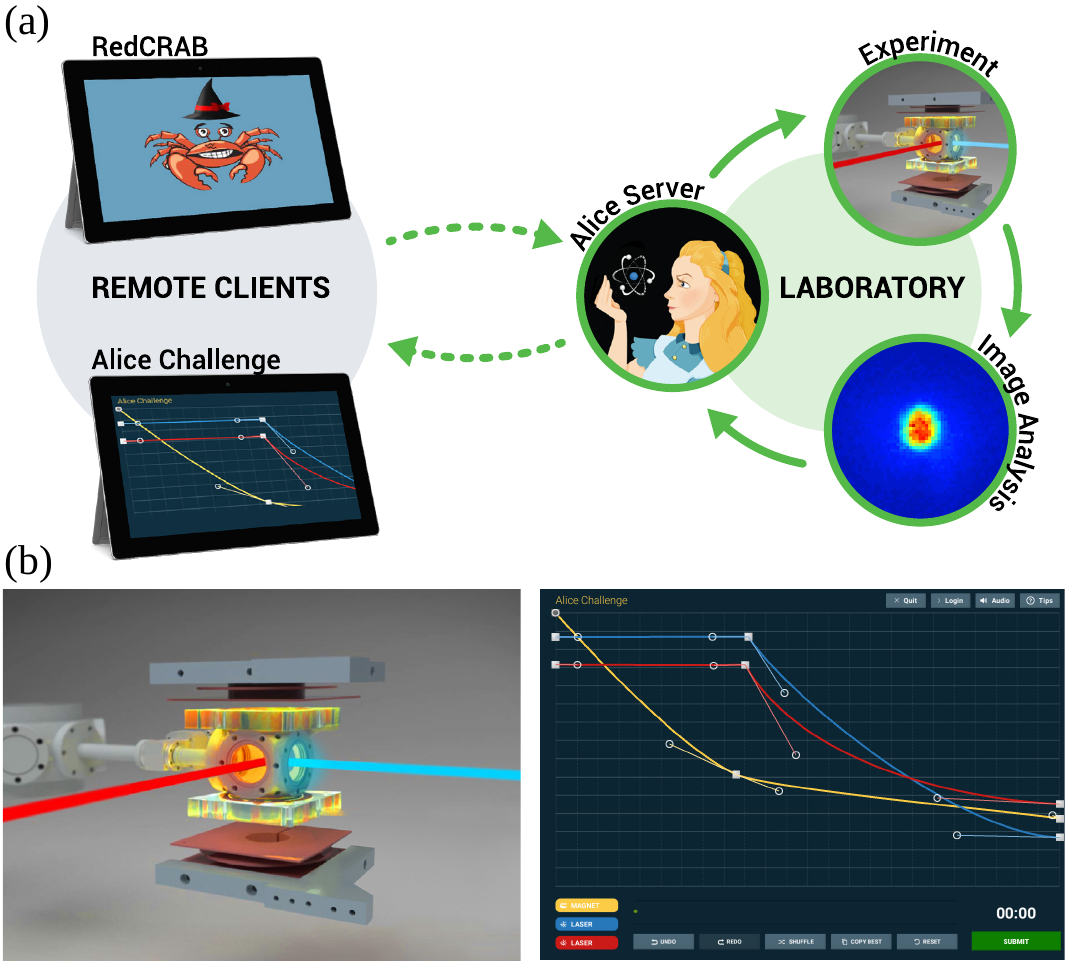}
    \caption{a)~Real-time remote scheme for RedCRAB and the Alice Challenge.
    The respective remote clients send experimental parameters through an online cloud interface which are turned into experimental sequences and executed by the Alice control program. The number of atoms in the Bose-Einstein condensate ($N_\mathrm{BEC}$) serves as a fitness value and is extracted from images of the atom cloud taken at the end of each sequence. The Alice control program closes the loop by sending the resulting $N_\mathrm{BEC}$ back to the remote clients through the same cloud interface. b)~Screenshots of the \href{http://alice.scienceathome.org/}{\emph{Alice Challenge}}~\cite{alicechallenge}. The left panel schematically shows the experimental setup. Players can control the magnetic field gradient depicted by the yellow shaded coils and the two dipole beams in red and blue. The control happens in the game client 
    (right panel) and features a spline editor for shaping the ramps for which the same color coding was used.} 
    \label{fig:overview_figure}
\end{figure}

The central purpose of this paper is to combine remote experimentation and citizen science with the aim of studying quantitatively how humans search while navigating the complex control landscape of Bose-Einstein condensate (BECs) production (figure~\ref{fig:overview_figure}a). Before presenting the results of this \href{http://alice.scienceathome.org/}{\textit{Alice Challenge}}~\cite{alicechallenge} we first characterize the landscape using a non-trivial heuristic that we derive from novel analysis of our previous citizen science work~\cite{Sorensen2016}.    

%-----------------------------------------------------
  % Figure Overview
%-----------------------------------------------------

\subsection*{Initial landscape investigations}
On \url{www.scienceathome.org}, our online citizen science platform, more than 250,000 people have so far contributed to the search for novel solutions to fast, 1D single atom transport in the computer game \textit{Quantum Moves}~\cite{Sorensen2016}. Surprisingly, clustering analysis of player solutions were found to bunch into distinct groups with clear underlying physical interpretations. We refer to such a group of related solutions as a strategy in the \emph{fitness landscape}: a function $J(\vec{u}) \in \mathbf{R}$ given by the control variables $\vec{u} \in \mathcal{S}$, where 
$\mathcal{S}$ is the set of possible solutions~\footnote{It is important to note that the topology of the fitness landscape critically depends on the choice of \emph{parametrization} or \emph{representation}.}. 
Here, we introduce the heuristics to explore the space between the identified strategies along the high-dimensional vector connecting the two as the {\em strategy connecting heuristic} (SCH). Perhaps surprisingly, we identify a narrow path of monotonously increasing high-fidelity solutions between the two strategies, which we denote a \emph{bridge} (see~\ref{SIA} for details). This demonstrates that for this problem a continuum of solutions with no clear physical interpretation can be traced out if all hundreds of control variables are changed synchronously in the appropriate way.

This leads us to ask whether established strategies in physics are truly distinct, or if they are simply labels we attach to different points in a continuum of possible solutions due to our inability to probe the entire solution space. In the latter case, coupled with the human desire to create identifiable patterns, this might cause us to terminate our search before discovering the true global optimum. This premature termination of search nicely illustrates the \emph{stopping problem} \cite{Freeman1983,blei_science_2017} considered in both computer algorithms and social science: determining criteria to stop searching when the best solution so far is deemed ``good enough''.
    
\begin{figure*}[tb]
	    \centering
	    \includegraphics[width=0.8\linewidth]{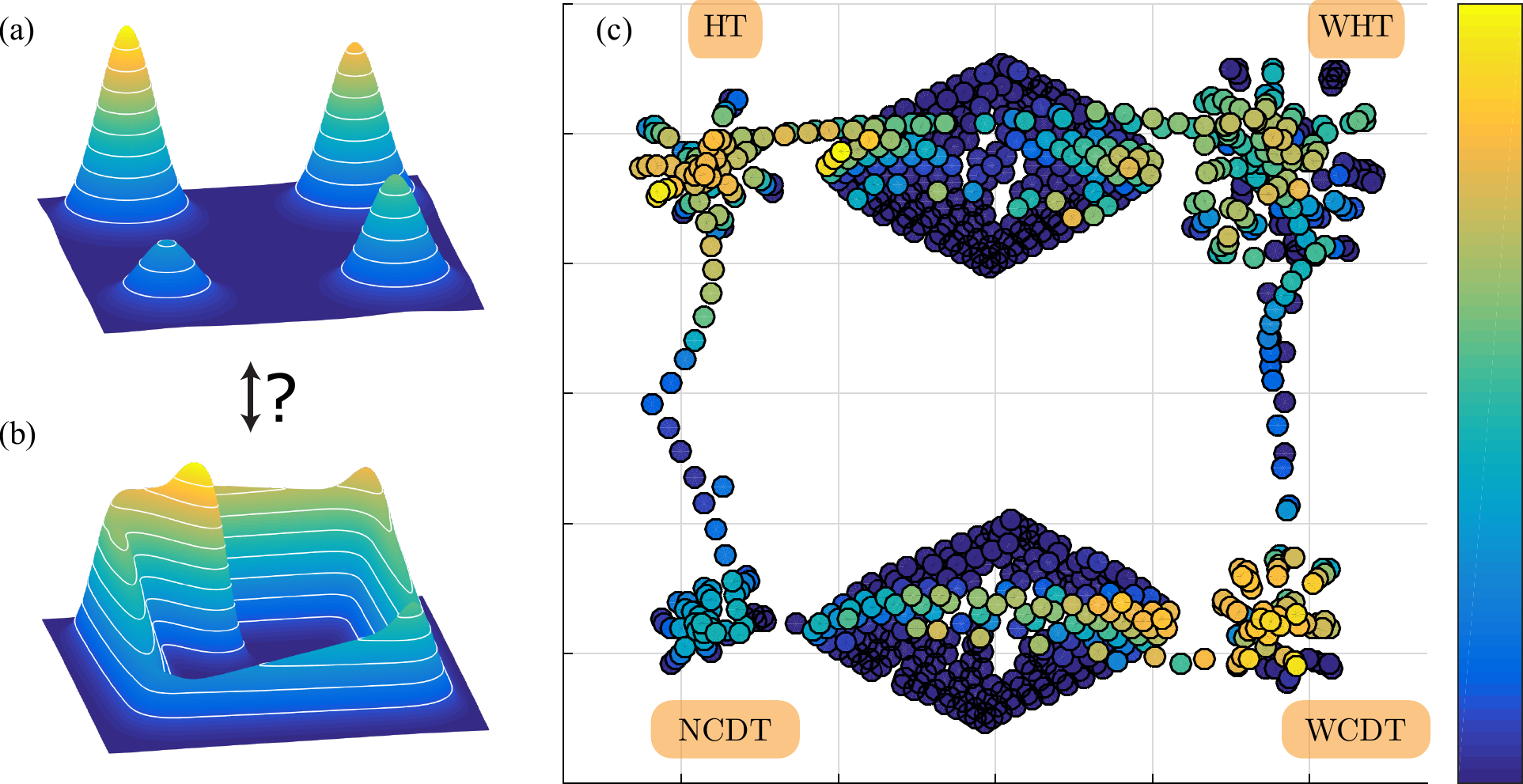}
		\caption{(a) Illustration of the apparent global landscape topology of distinct local optima after performing 1D parameter scans. However, as (b) illustrates, connecting \emph{bridges} were found both between some of the conventional strategies and to novel high-yield solutions in the high-dimensional search space. (c)~2D tSNE~\cite{vanderMaaten2008} representation of the landscape showing the variety of different trap configurations that are accessible in our experiment~\cite{footnote_tsne}. The plot contains the four main configurations which were scanned and optimized by 1D and 2D parameter scans. For more details, see text.}
		\label{fig:tsne}
\end{figure*}

We now apply this methodology to the high-dimensional problem of experimental BEC production~\cite{Ketterle1999}. In our case, increased BEC atom number will provide significantly improved initial conditions for subsequent quantum simulation experiments using optical lattices~\cite{Gross2017}. Although extensive optimization has been applied to the BEC creation problem over the past decade by employing global closed-loop optimization strategies using genetic algorithms~\cite{Rohringer2008,Rohringer2011,Geisel2013,Lausch2016}, little effort has been devoted to the characterization of the underlying landscape topology and thereby the fundamental difficulty level of the optimization problem. 
In the global landscape spanned by all possible controls, it is thus unknown if there is a convex optimization landscape with a single optimal strategy for BEC creation, individual distinct locally optimal strategies of varying quality (as illustrated in figure~\ref{fig:tsne}a), or a plethora of (possibly) connected solutions 
(figure~\ref{fig:tsne}b). Recent experiments~\cite{Wigley2016a} indicate that the underlying landscape is trap-free, however, this study did not explicitly optimize $N_\mathrm{BEC}$ and operated within a severely restricted subspace.  

In our experiment~\cite{Bason2016}, we capture \Rb atoms in a trap made of two orthogonal, focused \SI{1064}{\nano\meter} laser beams and a superimposed quadrupolar magnetic field which creates a magnetic field gradient at the position of the atoms and thereby forms a magnetic trap (see figure~\ref{fig:overview_figure}b for an illustration). We evaporatively cool the atoms past the phase transition to a BEC by lowering the intensity of the laser beams as well as the magnetic field gradient. Then, the traps are turned off, and the atoms are imaged 
with resonant light. Image analysis yields the total and condensed atom numbers $N_\mathrm{tot}$ and $N_\mathrm{BEC}$. 
    
This setting allows for evaporative cooling in two widely used trap configurations. First, making use of only the laser beams, a purely optical trap can be created, commonly known as a \textit{crossed dipole trap} (CDT)~\cite{Grimm1999}. Second, a single laser beam can be combined with a weak magnetic gradient to form a \textit{hybrid trap}~\cite{Lin2009}. In both cases, the traps are initially loaded from a pure tight magnetic trap (see \ref{SIB}). Conventionally, two types of geometrically differing loading schemes are pursued: loading into a large volume trap, which exhibits a nearly spatially mode-matched type of loading from the magnetic trap into the final trap configuration~\cite{Ketterle1999}, or a small volume trap with only a small spatial overlap. The latter leads to a ``dimple'' type  loading~\cite{Pinkse1997}  
in which a smaller but colder atom cloud is produced. We can directly control the effective volume of the trap by translating the focus position of one of the dipole trap beams. 
This inspired us to identify four initial ``conventional'' trap configurations (BEC creation strategies): a small volume, narrow crossed dipole trap (NCDT), a large volume, wide, counterpart (WCDT) and similarly a hybrid (HT) and wide hybrid trap (WHT). 
    
%-----------------------------------------------------
  % Figure TSNE plot 
%-----------------------------------------------------

We first optimize the system by applying typical researcher heuristics: starting from the set of control variables (see~\ref{SIB} for details) associated with a known strategy, we iteratively perform 1D scans of single variables  until a specified level of convergence is reached. 
The 1D scans yield four distinct solutions with the HT as the best performing strategy (see \ref{SIB}). This hints at the  landscape topology sketched in  figure~\ref{fig:tsne}a. 
Further systematic studies would then proceed to scans of two or more parameters simultaneously. However, allowing for scans of combined parameters enables prohibitively many different 2D parameter scans. Therefore, we proceed by applying the SCH derived from the Quantum Moves investigations. 

Both the low-yield  NCDT configuration and the WCDT are types of crossed dipole traps but with different effective volumes and thus represent ideal candidates for first exploration. However, a simple linear interpolation of all the available parameters between the NCDT and the WCDT fails to locate a bridge. 
Treating the effective trap volume as an independent second parameter 
realizes an extended 2D-interpolation and leads to the emergence of a bridge (see \ref{SIB}). 
In this case, the change of the trap depth induced by changing the trap volume has to be counterbalanced by a quadratic increase of the laser intensities involved.
Thus, changing to a different representation (\ie a particular combination of parameters) efficiently encapsulating the underlying physics yields a bridge and disproves the local character of the solution strategies involved. To illustrate this data, we created a dimensionality-reduced visualization~\cite{vanderMaaten2008} of the parameters scans (see figure~\ref{fig:tsne}c). The four initial strategies, investigated and optimized through the 1D scans, are represented by the four clusters in the corners. The data points forming the bridge between NCDT and WCDT lie in the diamond shape at the bottom. A few other 1D and 2D interpolations between other pairs of strategies are shown, but none is forming a bridge. 
In an attempt of locating a bridge between NCDT and HT extended 3D scans are performed (see \ref{SIB}, not shown in figure~\ref{fig:tsne}c). They identify a novel optimum away from the four initially defined experience-based trap configurations. This demonstrates that our initial candidate for a global optimum, the HT, is not a local optimum either when appropriate parameter sets are investigated. One is therefore inclined to view the topology of the landscape as closer to what is depicted in figure~\ref{fig:tsne}b, where the four conventional strategies are now connected with bridges and another (or many) higher-yield solutions exist in the full landscape.

Having established that the global optimum must be found outside the conventional strategies, 
we switch to the main topic of the paper: a remotely controlled strategy employing 
closed-loop optimization performed by experts and citizen scientists. As detailed below, our particular implementation allows for a 
quantitative assessment of the citizen scientist search behavior, but only a qualitative assessment of their absolute performance.
As a result, the search behavior of the two parties can only be compared qualitatively.

\subsection*{RedCRAB optimization}

As mentioned above, closed-loop optimization has been explored extensively for BEC creation using random, global 
methods~\cite{Rohringer2008,Rohringer2011,Geisel2013,Lausch2016,Wigley2016a} and is also routinely used to tailor radio-frequency fields to control nuclear spins or shape ultra-short laser pulses to influence molecular dynamics (see~\cite{Brif2010} and references therein). 
In our remote-expert collaboration, we employ the dCRAB~\cite{Rach2015} algorithm which is a basis-adaptive variant of the CRAB algorithm~\cite{Doria2011}. The main idea of both algorithms is to perform local landscape explorations, using control fields consisting of a truncated expansion in a suitable random basis. This approach makes optimization tractable by limiting the number of optimization parameters and has, at the same time, the advantage of obtaining information of the underlying landscape topology. It has been shown that the \textit{unconstrained} dCRAB algorithm converges to the global maximum of an underlying trap-free landscape with probability one~\cite{Rach2015}. That is, despite working in a truncated space, iterative random function basis changes allow the exploration of enough different directions in the functional space to escape traps induced by the reduced explored dimensionality~\cite{russell_control_2017,caneva_complexity_2014}. 
CRAB was introduced for the theoretical optimization of complex systems in which traditional optimal control theory could not be applied~\cite{Frank2016}. In closed-loop, CRAB was applied to optimize the superfluid to Mott insulator transition~\cite{Rosi2013}. Very recently dCRAB~\cite{Rach2015} was employed to realize autonomous calibration of single spin qubits~\cite{Frank2017} and to optimize atomic beam splitter sequences~\cite{Weidner2018}.

%-----------------------------------------------------
      % Figure RedCRAB/Alice results
    \begin{figure*}[t]
    \centering
\includegraphics[width=1\linewidth]{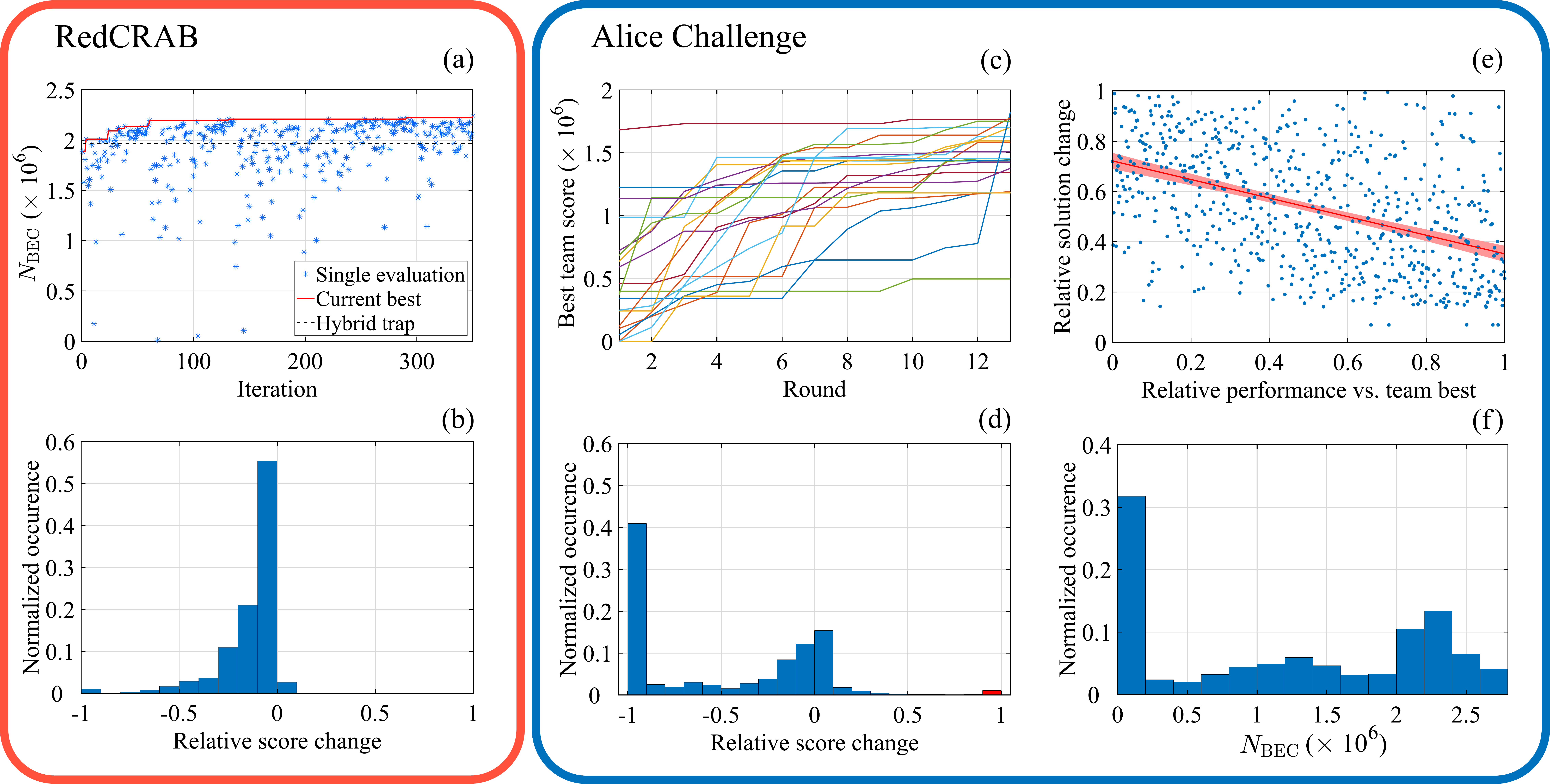}
    		\caption{(a)~Optimization with RedCRAB. $N_\mathrm{BEC}$ is plotted as a function of optimization algorithm iteration step. The blue data points indicate \emph{effective} single evaluations of $N_\mathrm{BEC}$ (see also \ref{SIC}). The red solid line denotes the current best $N_\mathrm{BEC}$. Comparing to the level of the HT (black dashed line), $N_\mathrm{BEC}$ was improved by \SI{20}{\percent}. (b)~Investigation of relative score changes compared to the current best solution for the RedCRAB optimization. (c)~Round-based best performance in the ATC. Each line shows the result for a team with three or more active players. Although human players had only a very limited number of tries (13 rounds), they still achieve relatively good optimization scores. Overall, all teams but one achieve scores above \num{1e6}. (d)~Investigation of relative score changes relative to the current best solution for the ATC.  The red bar summarizes all solutions which showed a relative score change $> 0.9$. (e)~All submissions in the ATC: How much do players edit their own solution compared with the relative team performance in the previous round. A linear regression with a \SI{95}{\percent} confidence bound is shown in red and yields a correlation of \num{-0.37(4)}. The distance measure is relative to players own previous solution. Both the distance and score measures are ranked within each round with the team-best score as a reference point.  (f)~Histogram for the achieved number of condensed atoms, $N_\mathrm{BEC}$, for all submitted solutions in the ASC. More than \SI{73}{\percent} of the submitted solutions were successful and yielded a BEC.}
    		    \label{fig:alice_redcrab_results}
    \end{figure*}
%%-----------------------------------------------------
    
Unlike closed-loop optimization performed in the past on other experiments, in which the optimization libraries were installed directly in the lab control software, we implemented the dCRAB remotely via the Alice remote interface \emph{RedCRAB} (see figure~\ref{fig:overview_figure}a). This gives the optimization experts direct feedback on the performance of the algorithm allowing them to apply real-time adaptations based on their previous experience as well as continually implementing algorithmic improvements to the control suite for future experiments. As in the case of the IBM Quantum Experience, we believe that this will  enhance the efficiency of experimentation as well as lower the barrier for even wider adoption of automated optimization in different quantum science and technologies aspects, from fundamental science experiments to technological and industrial developments. 

Here, the RedCRAB controls the intensity ramps of the two dipole trap beams, their duration $T_\mathrm{ramp}$ as well as a single parameter that represents the value of the magnetic field gradient during evaporation (see \ref{SIC}). As illustrated in figure~\ref{fig:alice_redcrab_results}a, we achieve a new maximal solution in about one hundred iterations, exceeding the result of the HT by more than \SI{20}{\percent}.

The solution is novel in the sense that it can be seen as a type of a CDT which is combined with the magnetic field gradient of the HT. The beam intensities are adjusted to lead to relatively similar trap depths as in the HT. However, especially in the beginning of the evaporation process, the trap is relaxed much faster leading to an overall shorter ramp. 
By applying SCH, a bridge connecting the HT to this novel solution could be identified (\ref{SIC}). 
This illustrates that the RedCRAB algorithm is highly effective both at locating novel, non-trivial optimal solutions as well as providing topological information of the underlying landscape. It substantiates the appearance of a complex but much more connected landscape than initially anticipated. 

\subsection*{Citizen science optimization}
In our second approach to remote optimization, we involve citizen scientists by employing a gamified remote user interface. We face the challenge of turning the adjustment of laser and magnetic field ramps into an interactive, engaging game. Therefore, we developed a client using the cross-platform engine \emph{Unity} and promoted it through our online community \url{www.scienceathome.org}. As depicted in figure~\ref{fig:overview_figure}b, the ramps are represented by three colored spline curves and are modified by adjustable control points. The total ramp duration, $T_\mathrm{ramp}$, is fixed to \SI{4}{\second}. After manipulating the splines, the player submits the solution which is then executed on the experiment in the lab (see figure~\ref{fig:overview_figure}a). The obtained $N_\mathrm{BEC}$ provides performance feedback to the players and is used to rank players in a high score list. Players can manipulate each of the splines as well as decide to see, copy and adapt everyone's previous solutions. This setup generates a collective search setting, where players emulate a multi-agent genetic search algorithm.

Citizen scientists have shown that they can solve highly complex natural science challenges~\cite{Sorensen2016,Cooper2010,Lee2014c,Kim2014}. However, data from previous projects suffer from the fact that they merely showed \emph{that} humans solve the challenges, but didn't answer \emph{how} a collective is able to balance local vs.\ global search while solving these complex problems. 
Social science studies in controlled lab settings have shown that individuals adapt their search based on performance feedback \cite{billinger2013search}. Specifically, if performance is improving, humans tend to make smaller changes (\ie local search), while if performance is worsening humans tend to make larger changes (\ie search with a global component). Therefore, experimental evidence suggests that human search strategies are neither purely local nor global \cite{billinger2013search, Vuculescu2017}.
Furthermore, studies have also established the importance of social learning and how humans tend to copy the best or most frequent solutions \cite{Mason2012, morgan2012evolutionary, muthukrishna2016and}, which facilitates an improved collective search performance. However, these laboratory-based studies have been constrained by the low dimensionality and artificial nature of the tasks to be solved. This raises concerns with respect to the external validity of the results: are these general human problem-solving patterns or are they merely behaviors elicited by the artificial task environment? Finally, previous citizen science results were based on intuitive game interfaces such as the close resemblance to sloshing water in Quantum Moves. In contrast, the Alice Challenge is not based on any obvious intuition. The question now is if and how citizen scientists would be able to efficiently balance local and global search when facing a real-world, rugged, \emph{non-intuitive} landscape?

In order to address this question, we created a controlled setting, the {\em Alice Team Challenge} (ATC). Unfortunately, due to the structure of the remote participation sufficient data could not be gathered to quantitatively study both the initial search behavior and the convergence properties of the human players (see \ref{SID}). Our previous work~\cite{Sorensen2016} demonstrated that the human contribution lay in roughly exploring the landscape and providing promising seeds for the subsequent, highly efficient numerical optimization. We therefore chose a design focusing on the initial explorative search of the players knowing that this would preclude any firm statements of the absolute performance of the players in terms of final atom number. Concretely, teams of five players each were formed, with every team member being allowed one submission in each of the thirteen rounds. After the five solutions from the active team were collected, they were run on the experiment and results provided to the players. Each round lasted about 180 seconds, thus a 13 round game lasted approximately one hour in total. 

As illustrated in figure~\ref{fig:alice_redcrab_results}c players showcase substantial, initial improvements across all game setups, which is evidence that humans can indeed effectively search complex, non-intuitive solution spaces (see also figure~\ref{SIfig7}). In order to make sense of how citizen scientists do this, we asked some of the top players how they perceived their own gameplay. One of them explained how he tried to draw on his previous experience as a microwave engineer when applying a black-box optimization approach. Because he apparently did not need a detailed understanding of the underlying principles of the search space, this suggests that humans might have domain generic search heuristics they rely on, when solving such high dimensional problems.

The player setup as well as differences in the accessible controls precludes a direct comparison of the absolute performance of the RedCRAB and the citizen scientists. However,  figures~\ref{fig:alice_redcrab_results}b and~\ref{fig:alice_redcrab_results}d demonstrating the distribution of the relative score changes clearly reveal how fundamentally different the respective search behaviors are. 
The local nature of the RedCRAB algorithm leads to incremental changes, either in positive or negative directions. \SI{80}{\percent} of the guesses differ by only \SI{20}{\percent} or less compared to the current best $N_\mathrm{BEC}$. 
In contrast, humans engage in many search attempts that lead to poor scores. Here, \SI{60}{\percent} of the solutions yield $N_\mathrm{BEC}$ which differs by \textit{more} than \SI{20}{\percent} compared to the current best.

To investigate this quantitatively, we further analyzed observations from 110 players in the ATC. Supporting previous lab studies \cite{billinger2013search, Vuculescu2017}, results show how players engaged in adaptive search; \ie if one had identified a good solution, compared to the solutions visible to the player, the player tended to make small adjustments in the next attempt. In contrast, if the player relatively speaking was far behind the best solution, the player tended to engage in more substantial adjustments to the current solution (see figure~\ref{fig:alice_redcrab_results}e and~\ref{SII}). Advancing previous studies, we were also able to isolate how players engaged in collective, adaptive search, \ie when they copied a solution from someone else in the team and subsequently manipulated it, before submitting the solution (\ref{SII}). The nature of adaptive search leads to a heterogeneous human search `algorithm' that combines local search with a global component. This search is prevented from stopping too early, since poorly performing individuals search more distantly, breaking free from exploitation boundaries, while individuals that are near the top perform exploitative, local search.  This out-of-the-lab quantitative characterization of citizen science search behavior represents the main result of this paper.

Finally, in order to  qualitatively explore the absolute performance of  the citizen scientists,  we created an open \emph{swarm} version of the game, the \emph{Alice Swarm Challenge} (ASC). The client was free to download for anyone and the number of submitted solutions was unrestricted. Participants could copy and modify other solutions freely. As this setting was uncontrolled, general statements about the search behavior are not possible.
In the ASC, we had roughly 500 citizen scientists spanning many countries and levels of education. 
The submitted solutions were queued, and an estimated process time was displayed. In this way, players could join, submit one or a set of solutions and come back at a later time to review the results. The game was open for participation for one week, 24 hours per day, with brief interruptions to resolve experimental problems. As an additional challenge, the game was restarted two to three times per day while changing $T_\mathrm{ramp}$ as well as the suggested start solutions. In the total 19 sessions, we covered a range from \SI{1.75}{\second} to \SI{8}{\second}. Discounting four failed sessions, 7577 solutions were submitted. Without the restriction of game rounds, players were able to further improve  the solutions. Figure~\ref{fig:alice_redcrab_results}f shows the distribution of the attained $N_\mathrm{BEC}$ across all sessions. For short ramp durations it became increasingly difficult to produce BECs with high $N_\mathrm{BEC}$ (see \ref{SIF}). Nonetheless the players could adapt to these changing conditions and produce optimized solutions.

The largest BEC was found for $T_\mathrm{ramp}=\SI{4}{\second}$ and contained about \SI{2.8e6}{\atoms}, which set a new record in our experiment. The solutions found by the players were qualitatively different from those found by numerical optimization. Where the RedCRAB algorithm was limited by only having control over the evaporation process and being able to apply only a single specific value for the magnetic field gradient, the players had full control over all ramps throughout the whole sequence of loading and evaporation. This was utilized to create a smoother transition from loading to evaporation. The magnetic field gradient during evaporation was initially kept at a constant value but relaxed towards the end (\ref{SIF}).

In conclusion, we have introduced a novel interface that allowed for the first remote closed-loop optimization of a BEC experiment, both with citizen scientists interacting through a gamified remote client and by connecting to numerical optimization experts. Both yielded solutions with improved performance compared to the previous best strategies. The obtained solutions were qualitatively different from well-known strategies conventionally pursued in the field. This hints at a possible continuum of efficient strategies for condensation. Although quantitative studies of player optimization performance was precluded by the design, it is striking that the players seemed to be able to compete on overall performance with the RedCRAB algorithm and also exhibited the ability to adapt to changes in the constraints (duration) and conditions (experimental drifts). The controlled design of the ATC yielded quantitative insight into the collective adaptive search performed by the players. This points toward a future in which the massive amounts of data on human problem solving from online citizen science games could be used as a resource for investigations of many ambitious  questions in social science.

\subsection*{Acknowledgements}
We thank J.\ Arlt for comments on the manuscript and valuable experimental input. SM gratefully acknowledges the support of the DFG via a Heisenberg fellowship and the TWITTER grant. The Aarhus team thanks the Danish Quantum Innovation Center, ERC, and the Carlsberg, John Templeton, and Lundbeck Foundations for financial support and National Instruments for development and testing support.

%\bibliography{library_strategies_in_physics,pnas-zotero}

%%%%%%%%%% Merge with supplemental materials %%%%%%%%%%
\balancecolsandclearpage
%\widetext
\begin{center}
\textbf{\large Supporting Information}
\end{center}
%%%%%%%%%% Merge with supplemental materials %%%%%%%%%%
%%%%%%%%%% Prefix a "S" to all equations, figures, tables and reset the counter %%%%%%%%%%
\setcounter{footnote}{0}
\setcounter{equation}{0}
\setcounter{figure}{0}
\setcounter{table}{0}
\setcounter{page}{1}
%\makeatletter
\renewcommand{\theequation}{S\arabic{equation}}
\renewcommand{\thefigure}{S\arabic{figure}}
\renewcommand{\thetable}{S\arabic{table}}
\renewcommand{\thesubsection}{SI,\,\Alph{subsection}}
\renewcommand{\bibnumfmt}[1]{[S#1]}
\renewcommand{\citenumfont}[1]{S#1}
%%%%%%%%%% Prefix a "S" to all equations, figures, tables and reset the counter %%%%%%%%%%
\subsection{Theoretical optimization of single particle transport}
\label{SIA}
In the following we briefly review the theoretical framework for gamified investigation of single atom transport in a controllable potential at the quantum speed limit (QSL)~\cite{Sorensen2016SI}. The framework is the citizen science game \emph{Quantum Moves}~\cite{lieberoth_getting_2014} and in particular the specific level \emph{Bring Home Water} (BHW). Here a graphically illustrated wavefunction of an atom in one dimension ($|\psi\rangle$) must be collected from a static Gaussian shaped potential well (optical tweezer) and subsequently transported into the ground state within a designated target area.  To realize this, the player dynamically adjusts the depth and position of a transport tweezer. The fraction of the state in the target state (the motional groundstate, $\ket{\psi_\text{T}}$) is given by the fidelity $0\leq F \leq1$ defined as $F=|\langle \psi_\text{T}|\psi\rangle|^2$. The player must also reach this state as quickly as possible (promoted by the introduction of a time penalty in the game).
Due to constraints in available resources, computer-based optimization at the QSL exhibits exponentially growing complexity~\cite{Sorensen2016SI,bukov_machine_2017}. Solutions to this particular challenge are valuable, for example for the realization of a large scale quantum computer based on ultracold atoms in optical lattices~\cite{Weitenberg2011a} and optical tweezers~\cite{Kaufman2015,Barredo2016,Bernien2017}.

%-----------------------------------------------------
  % Figure interpolating BHW solutions
\begin{figure*}[t]
\centering
\includegraphics[width=0.8\linewidth]{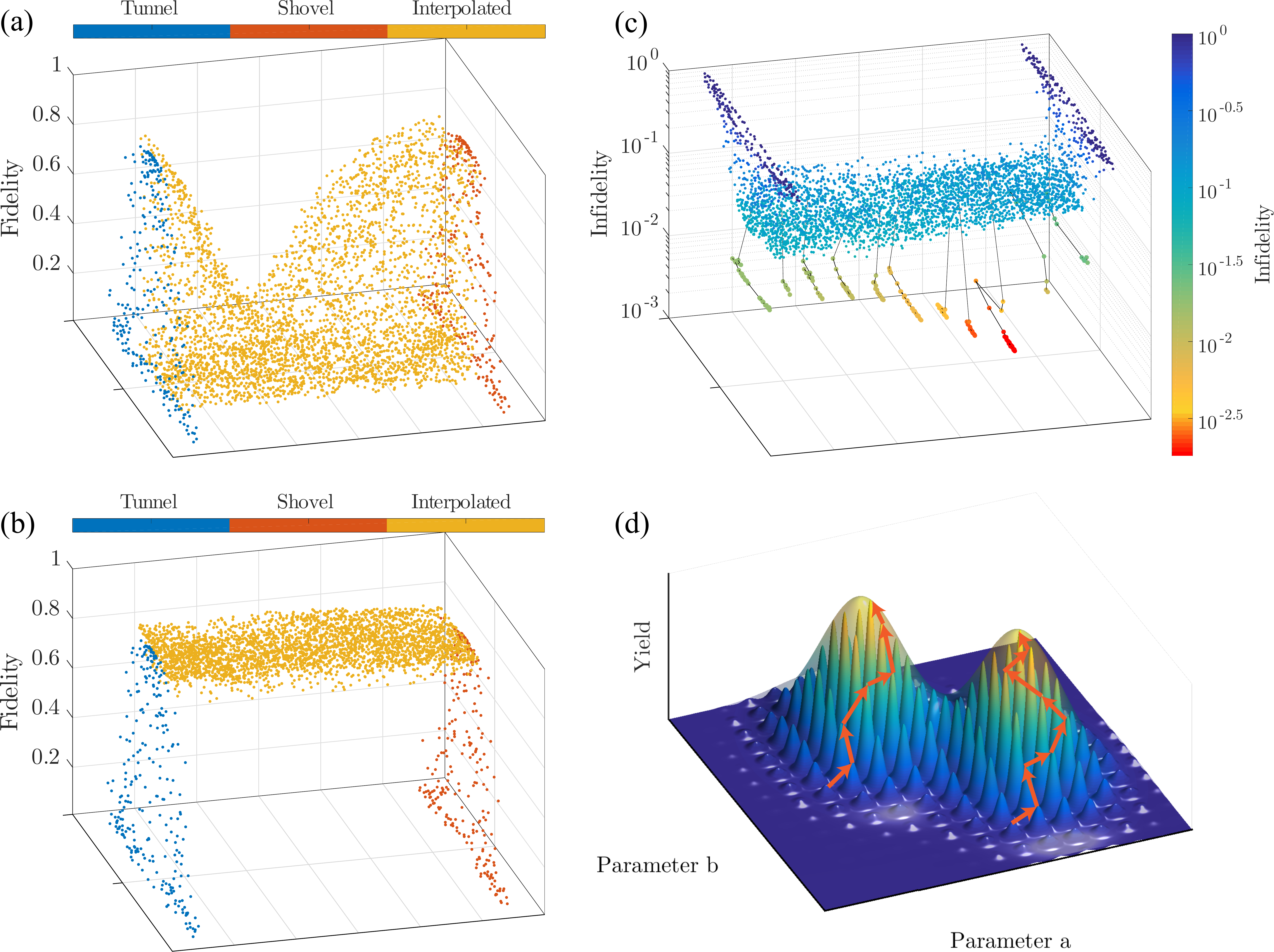}
		\caption{Visualization of the construction of a \emph{bridge} between the two clans of solutions of the BHW game. The process time was set to $T=0.19$ (for a definition of units see~\cite{Sorensen2016SI}), which is below the estimated QSL for both tunneling and shoveling clans ($T=0.25$ and $T=0.20$ respectively). The colors in (a) and (b) denote the type of each solution.  Note that for representation, the dimensionality has been reduced to two. Therefore, distances on the two horizontal axes should not be seen absolute and axes labeling was omitted. (a)~The result of using convex combinations of solutions and local perturbations of them to establish a connection between the two clans is shown. (b)~Local Krotov optimization was applied to the yellow marked points in (a) and close-to-optimal solutions are attained. (c)~Infidelity of resulting optimized solutions. Starting from the established bridge from (a), individual points were optimized using the GOLSS scheme (black lines). (d)~Illustration of the concept of the \emph{superlandscape}. Whereas the underlying optimization landscape consists of densely lying local optima, the superlandscape is defined as the smooth envelope function spanned on top of them.
    The two orange paths visualize the optimization with the GOLSS scheme by moving along the local optima of the underlying landscape towards an extremum in the superlandscape.}
	 		\label{fig:BHW_Krotov_collage}
\end{figure*}
%-----------------------------------------------------

The focus in quantum optimal control has been primarily on developing tailored local optimization algorithms like Krotov, GRAPE and CRAB~\cite{Sklarz2002, schirmer2011efficient,khaneja2005optimal, Caneva2011a}. The former two methods are very efficient, since they exploit the structure of the Schrödinger equation, whereas CRAB is universally applicable since it can use a gradient-free method to reach the optimum, and furthermore it has the attractive feature of operating in a reduced sized basis. Recently, gradient-based optimization in a reduced basis has also been exploited in the GOAT and GROUP algorithms~\cite{machnes_gradient_2015, Sorensen2018}. All these local methods are typically turned into global optimizers by restarting over a wide range of initial seeds until they give sufficiently good results.
In alternative efforts, global search methods such as Differential Evolution, CMA-ES, and reinforcement learning have been applied directly to quantum control~\cite{zahedinejad2014evolutionary,shir2011evolutionary,bukov_machine_2017} and very recently the local and global methods have been combined \cite{Sorensen2018}.

In Ref.~\cite{Sorensen2016SI} it was shown that the optimization of player solutions from the BHW challenge outperforms such purely numerical approaches for transport durations close to the QSL. The best results were found by optimizing the player solutions using the Krotov algorithm in a hybrid Computer-Human Optimization (CHOP). In order to compare the different solutions obtained with CHOP, a distance measure was introduced. 
A clustering analysis revealed that solutions fall into two distinct clusters denoted as ``clans''.
The solutions forming a clan all follow a similar strategy, to which one can assign a physical interpretation.  One of the clans exploits dynamics reminiscent of quantum tunneling, while the members of the other clan use a classically inspired  shoveling strategy~\cite{Sorensen2016SI}.

Here, we ask if these clans really represent physically distinct strategies in the sense that no mixed-strategy, high-yield solutions exist. Proving that a given solution is locally optimal and thus truly distinct involves extensive numerical work in either random sampling combined with methods like principal component analysis 
or systematic reconstruction of the full Hessian in the surrounding high-dimensional space~\cite{shir_efficient_2014}. Given the high dimensionality of the problem, an exhaustive exploration of the whole space is impossible to realize with a reasonable amount of time and resources. Instead, we investigated the topology of the landscape spanned by linear interpolation between the individual controls of representatives of the two clans. Given the interpolation parameter $\alpha\in[0,1]$, the interpolated control is defined as 
\begin{align}
     \vec{u}_\mathrm{int}(\alpha) = \alpha\,\vec{u}_1 + (1-\alpha)\,\vec{u}_2,
     \label{eq:interpolation}
\end{align}
where $\vec{u}_1$ and  $\vec{u}_2$ are the controls of the two solutions interpolated between. 
Figure~\ref{fig:BHW_Krotov_collage}a depicts a 2D visualization of the landscape corresponding to the interpolated solutions and local random perturbations to these using the t-SNE algorithm~\cite{VanDerMaaten2008} as applied in Ref.~\cite{Sorensen2016SI}. The rapid decline in fidelity of the interpolated points and the multitude of points yielding zero fidelity suggests that the clans can be seen as distinct regions of good, nearly optimal solutions in the underlying optimization landscape.

According to this interpretation, one would expect local optimization of these solutions to drag them towards either the shoveling or the tunneling solutions and thereby yield a region of attraction for each clan. 
Instead, local optimization using the Krotov algorithm  results in the high fidelity \emph{bridge} shown in figure~\ref{fig:BHW_Krotov_collage}b. 

In Ref.~\cite{Sorensen2016SI}, we introduce a distance map $D_{ij}$ for two solutions $i$ and $j$, which compares the overlap between two corresponding wavefunctions $\ket{\psi_i(x,t)}$ and $\ket{\psi_j(x,t)}$ at each time step $t$ for a given total transport time $T$:
\begin{align}
    D_{ij} = \frac{1}{T}\int^T_0 \langle f_{ij}|f_{ij} \rangle \text{d}t,\label{eq:distance}
\end{align}
where $ \ket{f_{ij}(x,t)} = \ket{\psi_i(x,t)} - \ket{\psi_j(x,t)} $ is the difference between the wave functions at each position $x$.

In terms of this distance metric, the displacement of each numerically optimized solution from the initial seed is relatively small: the optimization of the points in figure~\ref{fig:BHW_Krotov_collage}a leads to nearly vertical lines in the visualized landscape. This implies that as long as the right region of the landscape is explored, very close to the non-perfect trial solution lies a better solution. Additionally, each initial seed converged to a different optimum, i.\,e., new, distinct solutions have been found as illustrated by the yellow points of figure~\ref{fig:BHW_Krotov_collage}b. Thus, the landscape is locally very rugged, but rich in optima.

Given the density of locally optimal points, we now define the \emph{superlandscape} as the approximately smooth envelope function spanned by the optimal  points. If $\mathcal{O}$ is a local optimization algorithm then this landscape is the composition $\tilde{J}=J[\mathcal{O}[\vec{u}(t)]]$. 
Figure~\ref{fig:BHW_Krotov_collage}d illustrates a simple, generic superlandscape. The underlying landscape consists of a dense collection of individual peaks with smoothly varying heights. 

Physically, we interpret the sharpness of the peaks and the density of optima in terms of the population along the different instantaneous eigenvalues  of the problem: Due to the time-energy uncertainty relation, rapid transfer in the vicinity of the QSL can only be achieved by significant excitations, which ensure rapid relative phase evolution of the individual wave function components~\cite{Mandelstam1991}.  Towards the end of the process, the population needs to be refocused into the ground state. Any minor perturbation to a path at some instant will in general lead to a decreased fidelity. However, it can (nearly) be compensated with another carefully chosen perturbation at another time, leading to many closely spaced locally optimal solutions.

If the superlandscape can be evaluated with sufficient speed (using efficient local optimization), then we propose to  perform completely deterministic global optimization using a method that we call Gradient-Optimization and Local Superlandscape Search (GOLSS). Such a hybrid local-local optimization scheme has been recently proposed in the context of trap-free landscapes in the presence of noise~\cite{egger_adaptive_2014}. In contrast, GOLSS entirely eliminates the random steps normally present in global optimization and could therefore potentially offer significant speedup compared to existing methods. (See figure~\ref{fig:BHW_Krotov_collage}d for an illustration.)

Here, we use a Nelder-Mead type search \cite{Nelder1965} combined with Krotov optimization to implement GOLSS. We start this optimization of $\tilde{J}$ at a number of interpolated solutions along the identified bridge. This results in significantly improved solutions as shown in figure~\ref{fig:BHW_Krotov_collage}c. These solutions are found at a duration of $T=0.19$.  
The best optimized solutions from this combined search reached $F=0.998$ in fidelity, which is an improvement of nearly two orders of magnitude in terms of infidelity over the best player optimized solution (CHOP) yielding  $F=0.929$. 
It also represents an  improvement of the previously obtained numerical estimates of the QSL for both the tunneling and shoveling clans, which were at T=0.25 and T=0.20 respectively. When we inspect the actual solution it is clearly seen to be a combination of the tunneling and shoveling strategies, since it places the transport tweezer on top of the atom rather than to the left or right of it.

The fact that a single 1D-line scan identifies a bridge is an illustration of a deeper underlying principle in numerical optimization: whereas nearly all of the many possible search directions yield poor behavior (illustrated by the blue and red points in figure~\ref{fig:BHW_Krotov_collage}a), once a good heuristic encapsulating the essence of the problem is determined, low dimensional search is sufficient~\cite{newell_human_1972}.
Previously, in Ref.~\cite{Sorensen2016SI}, we constructed these search spaces explicitly using parametrizations that emerged from data analysis of large amounts of numerical solutions. Recently, a similar approach was applied to extract low-dimensional search spaces for spin-chain dynamics using ML-generated data~\cite{bukov_machine_2017}. The search along convex linear combinations of existing solutions introduced here may provide a computationally inexpensive methodology to identify promising search directions in the multi-dimensional landscape.  In addition, we believe that the concept of superlandscape and local search within it will be a useful metaheuristic for finding high-quality solutions for quantum optimal control problems. 

\subsection{Experimental details -- Parameter scans}
\label{SIB}
Each trap configuration that is presented in the main text is loaded from a pre-cooled \Rb atom cloud prepared in the $\ket{F=2,m_F=2}$ state and trapped in a magnetic quadrupole trap. At this stage, we typically have \SI{5e8}{\atoms} at a temperature of $\approx\SI{30}{\micro\kelvin}$. The experiments of the Alice Challenge start from this point. The CDT consists of two perpendicular beams which overlap in the horizontal plane. They have $1/e^2$ waists of \SI{45}{\micro\meter} (beam A) and \SI{85}{\micro\meter} (beam B), respectively. The longitudinal focus position $x_\mathrm{focus}$ of beam A can be adjusted, thereby changing its effective waist at the crossing point of the beams. This beam is used to realize HT and WHT. The beams are placed with a vertical offset of around \SI{90}{\micro\meter} below the centre of the magnetic trap. An offset magnetic field $B_\mathrm{off}$ in that direction can be used to tune this distance. 

In our experiment, the control of the time-dependent light and magnetic fields during the loading and evaporation is initially limited to eight parameters. 
Later, for the remote-controlled experiments we will relax this restricted representation to allow for the full high dimensional quasi-continuous control only restricted by hardware limitations

For the parameter scans, the intensity ramps $I(t)$ of the dipole trap beams are described by a function inspired by a simple model of evaporative cooling based on scaling laws~\cite{OHara2001a} 
\begin{align}
     \frac{I(t)}{I_i}=\left(1+\frac{t}{\tau}\right)^{-\beta}.
     \label{eq:Ohara_evaporation}
\end{align}
Here, $I_i$ is the initial intensity, whereas $\tau$ and $\beta$ influence the shape of the ramp. The duration of the ramp is fixed by defining the ratio of initial and final intensity $I_i/I_f$ for a given $\tau$ and $\beta$. For simplicity, the intensity ratio, as well as $\tau$ and $\beta$ are chosen to be the same for the two beams. $I_i$, however, is an independent parameter.

For the loading process from the magnetic trap into the final trap configuration, the dipole trap beams are regulated to their individual $I_i$ and the magnetic field gradient is lowered  in three linear ramps from \SI{130}{\gauss\per\centi\meter} initially to a final value $B'_f$, which is retained throughout the evaporation. In total, this leads to eight individual optimization parameters.

%-----------------------------------------------------
  % Figure 1D and 2D scans
\begin{figure}[htb]
	    \centering
	    \includegraphics[width=0.7\linewidth]{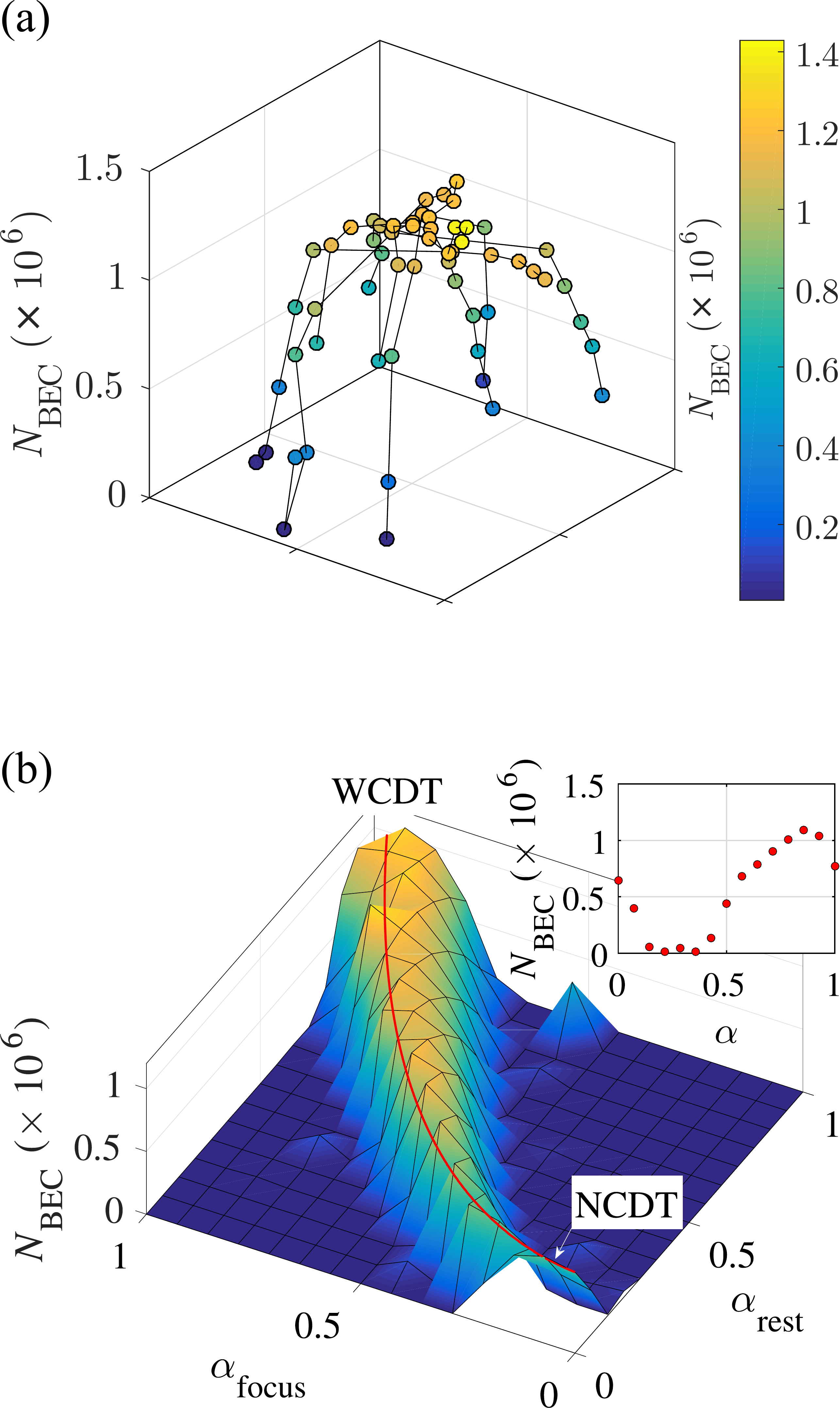}
		\caption{(a) 1D parameter scans used to find locally optimized parameters displayed in a 2D tSNE representation exemplarily shown for the hybrid trap. In this case, there is no indication for a connection to any of the other ``conventional'' trap configurations. (b)~2D-interpolation between NCDT and WCDT showing a connecting bridge between the two. Along the first dimension, the interpolation parameter $\alpha_\mathrm{focus}$ is varied influencing $x_\mathrm{focus}$. The remaining parameters are scanned synchronously along the second dimension described by the interpolation parameter $\alpha_\mathrm{rest}$. The solid red line is a simple theoretical estimate of where to find the bridge within the given parameter space. The inset shows a diagonal cut through the landscape and illustrates, a simple linear interpolation with all parameters fails to find a bridge.}
		\label{fig:tsne2}
\end{figure}
%-----------------------------------------------------

The result of 1D parameter scans is shown exemplary in figure~\ref{fig:tsne2}a for the case of the HT. The scans clearly reveal a peak-like structure with a set of \emph{1D-optimal} solutions. 
For the  conventional trap configurations (NCDT, WCDT, HT, WHT) we obtain the 1D-optimized values $N_\mathrm{BEC}=(\num{0.53(9)},\ \num{1.07(5)},\ \num{1.8(2)},\ \num{1.1(4)})\cdot10^6$~\cite{footnote_error}. 
The resulting duration $T_\mathrm{ramp}$ of the whole evaporation (counting from the beginning of the loading process) differed for the individual traps and reached $T_\mathrm{ramp} = (2.66,2.97,5.56,6.60)\si{\second}$, respectively. 

To investigate the topology of the landscape we search for interconnecting bridges by simultaneously scanning several parameters. 
Both  the low-yield  NCDT configuration and the WCDT are types of crossed dipole traps but with different effective volumes dictated by $x_\mathrm{focus}$. A simple linear interpolation of all the available parameters between the NCDT and the WCDT (\cf equation~\ref{eq:interpolation}) fails to locate a bridge as illustrated in the inset of figure~\ref{fig:tsne2}b. This is consistent with the BHW case treated above in which the bridge did not appear until local optimization was performed on the interpolated seeds. Since local optimization is very time consuming in the experimental case, we instead try to extend the search space slightly beyond the simple 1D case.

Treating $x_\mathrm{focus}$ independently and introducing a second interpolation parameter 
realizes an extended 2D-interpolation, which  leads to the emergence of a bridge as shown in figure~\ref{fig:tsne2}b. The necessity of a 2D scan can be interpreted in terms of the distinction Simon and Newell~\cite{newell_human_1972} make between problem space (the subjective search space) and the objective task environment (physical subprocesses): the chosen parametrization in terms of different laser beam and magnetic field settings versus the time dependent trap depths and shapes during the loading and evaporation processes. In this case, the change of the trap depth induced by changing the trap volume has to be counterbalanced by a quadratic increase of the laser intensities involved.  The solid red line in figure~\ref{fig:tsne2}b marks the position, which yields the same trap depth at the beginning of the  evaporation stage. 

Thus, changing to a different representation (\ie a certain combination of parameters) efficiently encapsulating the underlying physics  yields a bridge and disproves the local character of the solution strategies involved. 
We stress that although the bridge was located in a seemingly simple 2D scan, the heuristics introduced for the BHW case of identifying the multi-dimensional search direction  between  established strategies was crucial.

Comparing all other pairs of conventional strategies, the choice of parameter combinations for  extended interpolation is much harder to motivate physically and simple 1D- and 2D-interpolation fail to locate bridges.  

%-----------------------------------------------------
  % Figure 3D scan
\begin{figure*}[tb]
\centering
\includegraphics[width=1\linewidth]{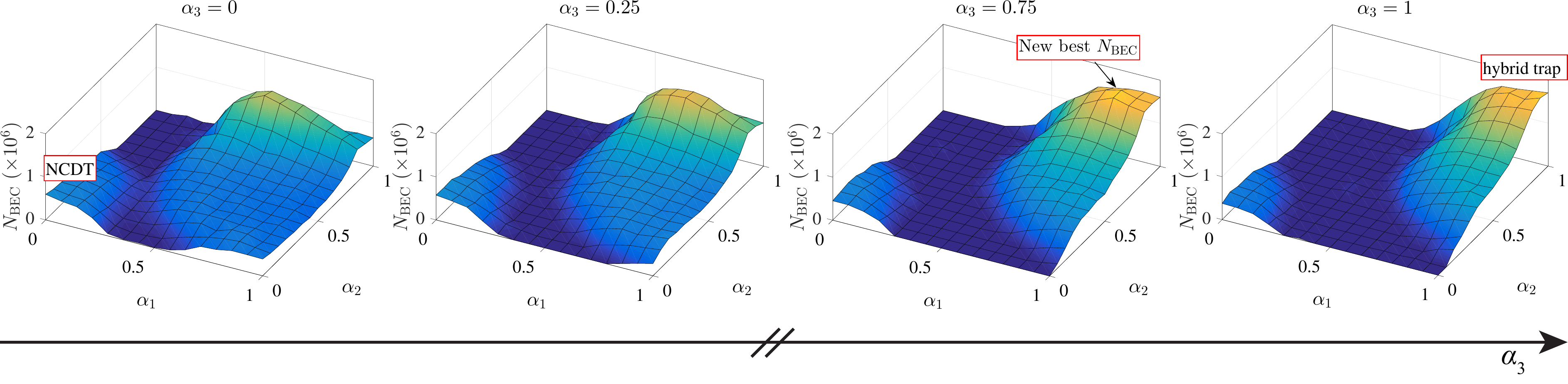}
\caption{Parameter scan in three dimensions from the NCDT to the HT. The normalized interpolation parameters are denoted $\alpha_{1,2,3}$ Each frame represents a scan point in the third dimension ($\alpha_3$). In frame $\alpha_3=0.75$, a trap configuration with $N_\mathrm{BEC}$ larger than for the HT is found. This shows that the HT is not a local optimum which is in contrast to the initial indications (see main text).}
\label{fig:3Dscan}
\end{figure*}
%-----------------------------------------------------

As described in the main text, parameter scans were not only performed in 1D or 2D. In figure~\ref{fig:3Dscan} the extension of the scan space in the third dimension is presented for parameter sets linearly interpolated between the NCDT and HT. An optimum that moves from frame to frame is revealed and a parameter set is found yielding a slightly higher $N_\mathrm{BEC}$ than in the HT is found. The scan disproves the local character of the HT which was implied by the 1D parameter scans.

\subsection{Experimental details -- Remote optimization}
\label{SIC}
For this investigation the search space is restricted to the domain between HT and NCDT by fixing the effective volume of our CDT.  Likewise, the vertical offset magnetic field is fixed to a value compensating the residual background magnetic fields and is not changed during optimization.
For both the remote optimization with RedCRAB and the Alice Challenge, non-optimized configurations were chosen as starting points of the optimization runs. The RedCRAB starting point was close to the HT configuration. In the Alice Challenge at each start of a new round, the high score list was emptied and filled with  low quality solutions yielding typically $N_\mathrm{BEC}\approx (1 - 2)\cdot 10^5$ atoms.

In contrast to the previous parametrization of the shape of the laser ramps, RedCRAB (based on the dCRAB algorithm~\cite{Rach2015SI}) focuses on a finite set of relevant basis functions that make up a sufficiently good ramp. Here, each of the ramps is composed of a Fourier basis up to the 5\textsuperscript{th} harmonics in units of $2 \pi / T_\mathrm{ramp}$, where $T_\mathrm{ramp}$ is the total ramp duration. $T_\mathrm{ramp}$ itself as well as $B'_f$ during evaporation  are chosen to be subject to optimization. The loading procedure of a certain trap configuration is the same predefined sequence described for the parameter scans above. To overcome shot-to-shot fluctuations and thus resulting in an optimization driven and influenced by noise, an adaptive averaging scheme is applied with a stepwise increasing number of averages for higher yields in $N_\mathrm{BEC}$. Outliers to high $N_\mathrm{BEC}$ are in this way re-evaluated. However, we still keep the number of time-consuming evaluations low at early stages of the optimization which decreases the overall convergence time.

The adaptive averaging scheme is implemented the following way: The current best $N_\mathrm{BEC}$ is denoted by $N_{\text{rec}}$. Based on this value, a set of threshold values 
(thv) are defined via \{thv1,thv2,thv3,thv4\} = 
\{$0.9N_{\text{rec}}$,$0.96N_{\text{rec}}$,$0.99N_{\text{rec}}$,$1.01N_{\text{rec}}$\}. Say the experimental apparatus 
will return the value $N_{\text{trial}}$ for a newly evaluated set of pulses. This value will have to go through the 
cascade of threshold values \{thv1,thv2,thv3,thv4\} before being chosen as the new current record $N_{\text{rec}}$. At 
first, $N_{\text{trial}}$ is compared to thv1. Only if it exceeds thv1, the same pulse will be re-evaluated yielding 
$\bar{N}_{\text{trial}}$ = $0.5(N_{\text{trail}}^{i-1} + N_{\text{trial}}^i)$. As a second step, 
$\bar{N}_{\text{trial}}$ is compared to thv2. Again, only if it exceeds thv2, the very same pulse is re-evaluated for a 
third time. This procedure is repeated as long as $\bar{N}_{\text{trial}}$ succeeds to jump over the respective thvn 
value until eventually $N_{\text{rec}}$ will be updated to $\bar{N}_{\text{trial}}$ after 5 successful re-evaluations.

The optimized solution that was found with RedCRAB is a mixture between NCDT and HT. $B'_f$  resembles closely the one of HT,  however \SI{25}{\percent} less intensity is used in beam A. This is partially compensated by adding beam B. This results in a trap depth which is about \SI{15}{\percent} lower compared to the HT at the beginning of the evaporation. The lowering rate of the trap is comparable in the first part of the evaporation and drops at the end below the one of the HT. At the same time the calculated geometric mean of the trap frequencies ($\bar{\omega}$) is higher. In both cases, the evaporation ends at a similar trap depth and similar $\bar{\omega}$. The total ramp duration is with $T_\mathrm{ramp}=\SI{4.92}{\second}$ shorter than the one of HT. 

\begin{figure*}[htbp]
	\centering
	\includegraphics[width=1\linewidth]{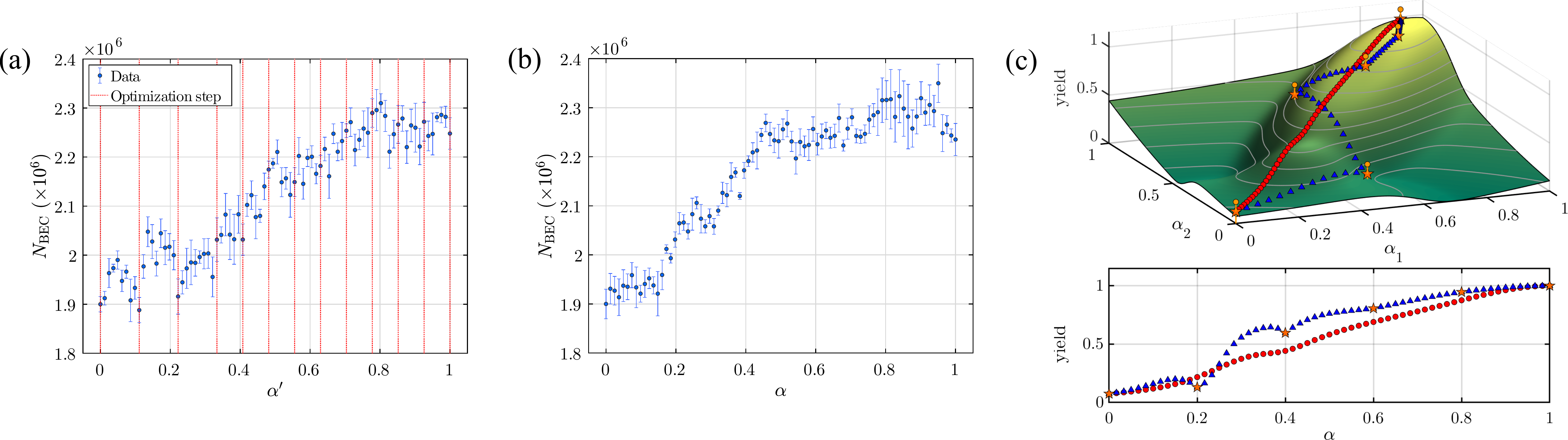}
	\caption{(a)~Stepwise, linear interpolation from the HT (first vertical dashed line) through the starting point of RedCRAB optimization (second vertical dashed line) to its optimum (last vertical dashed line). That is, the ramps between each optimization step (remaining vertical dashed lines) are interpolated. $\alpha'$ is a generalized parameter incorporating the individual interpolations. (b)~Direct linear interpolation between the ramps for the HT and the found optimum of RedCRAB optimization (figure~\ref{fig:alice_redcrab_results}a in main text). $\alpha$ is the interpolation parameter. The error bars represent the standard error for five repetitions. (c)~A possible way to explain the different results of (a) and (b) on an exemplified optimization landscape spanned by parameters $\alpha_1$ and $\alpha_2$. In the presence of noise, an optimizer like the RedCRAB algorithm could follow the non-ideal route towards the optimum marked by the orange triangles with error bars. However, following that trace by piece-wise interpolation (blue triangles, corresponding to data in (a)), renders the topology of the landscape visible and results in a non-monotonic trace. Direct linear interpolation (red circles, corresponding to data in (b)), in this case, exhibits continually increasing yields.}
	\label{fig:illustration_interpolation}
\end{figure*}

In the process of reaching this solution, RedCRAB identified of the order of 10 intermediate improved solutions (see steps of the current best $N_\mathrm{BEC}$ depicted as red solid line in main text figure~\ref{fig:alice_redcrab_results}a). If the underlying landscape is sufficiently smooth, one might expect to be able to locate a bridge between the standard strategies and the novel optimum by a linear or possibly non-linear combination of these intermediate solutions. As illustrated in figure~\ref{fig:illustration_interpolation}a, in which step-wise linear interpolation between these intermediate solutions is performed, this is nearly but not exactly the case. There are small intervals of decreasing yield. However, a direct linear interpolation between the HT and the novel optimum  yields a monotonically increasing bridge (see figure~\ref{fig:illustration_interpolation}b). Thus, the non-monotonicity of the stepwise interpolation is not likely to be caused by ruggedness of the underlying landscape. Rather, a more natural explanation would be that since the local simplex-optimization component of dCRAB is not purely gradient-based it does not necessarily have its axes oriented along the maximal slope and will have a tendency to find a slightly wiggly path towards the optimum. This increases the chance that experimental noise will occasionally cause the algorithm to find false search directions from which it is slowly recovering in the following iterations. Figure~\ref{fig:illustration_interpolation}c gives a graphical visualization of this phenomenon.

%-----------------------------------------------------
  % Figure interpolation illustration
%-----------------------------------------------------

\subsection{Design considerations of the Alice Challenge}
\label{SID}
Around the same time as the development of the remote interface for establishing a connection with RedCRAB, a simplified remote client with a graphical user interface was developed. It allows one to control the dipole beam intensities and the magnetic field gradient of the experiment via piecewise defined functions. Tests with an undergraduate student and a collaborator situated in the UK were successful. Both were allowed to optimize evaporation sequences via this client independently. Improved solutions were found which lead to the idea of gamifying the task of optimizing the evaporation process and give ``non-experts'' real-time access to our experiment.

The development of the Alice Challenge began. In the design phase, it was very unclear if large groups of citizen scientists could be recruited given the non-intuitive and relatively low level of gamificiation compared to previous games. To make recruitment more plausible we decided to run the challenge in a well-defined and relatively short time interval to ensure that we would have a fairly high number of simultaneous users at all times. A prototype version of the game was developed and presented at the National Instruments NIWeek 2016 in Austin, Texas. The event was used for a test-run of the game interface and to acquire a broad potential user base for the actual Alice Challenge.

Given the cycle time of the experiment, the relatively short duration of the Alice Challenge put severe restrictions on the total amount of data that could be collected. Due to the restricted amount of data available, we had to make two crucial choices. First, we had to decide between the controlled setting of teams optimizing in parallel or the full uncontrolled free access of every player to all previous optimization results. The former would allow for systematic investigation of the initial search behavior at the cost of obtaining little or no information about the convergence properties of the amateur players because each team would not have enough iterations to converge. 

Since participants in the ATC are not confined to a lab setting nor paid, dropouts can create a major obstacle for such team-based experiments. In abstract terms, there is a \((1-E)^N\) probability that all team members will finish all rounds, where $E$ is the dropout probability and $N$ is the number of people in the team. The bigger the $N$ and the more rounds, the bigger the dropout. Fewer and larger teams would also lead to fewer independent optimization runs and would therefore severely limit the statistical power of the quantitative characterization of player search behavior. This led us to settle on teams of five. Likewise, we settled on a relatively short amount of 13 rounds. Every team member was allowed one submission in each of the 13 rounds via the remote interface (see figure~\ref{fig:overview_figure}b in the main text). We explicitly instructed participants that they would be working together in a team, but that collaboration was possible only through the visibility of team members’ solutions and scores. \emph{After} the five solutions from  the active team were collected, they were run on the experiment and results provided to the players. Each round lasted about 180 seconds. Therefore, a 13 round game lasted approximately one hour in total. This prioritization is in line with the findings from our previous citizen science work: the strength of the algorithms is to optimize a given seed systematically, whereas the contribution of the players was to provide good seeds by a more global rough search of the landscape.

To further increase engagement of the participants, we emphasized the importance of finishing the game as well as increasing the resilience of the team experience by replacing dropouts with bots. Following a recruitment campaign based on snow-balling, 142 participants from around the world committed to taking part in the experiment. Participants selected up to 10 one-hour slots during the week of the challenge. Once the recruitment campaign was over, they were randomly assigned to teams, while maximizing the number of complete teams. In incomplete teams, the slots of the missing players were taken by computational agents who would simply reshuffle existing solutions. One game session had to be excluded from the dataset, because the experiment setup had drifted significantly and thus the evaluation function had been corrupted. Overall, due to no-shows, dropouts, as well as the corrupted session, we analyzed observations from 110 players out of the original 142. All participants gave explicit consent to participate in the study, which was approved by an IRB at Aarhus University.

We are aware that, whereas this design optimizes statistical power to investigate the initial explorative behavior of the players, it unfortunately did not leave much room for investigating how well the players performed in absolute terms. To slightly compensate for this, we also inserted a short, uncontrolled ``Swarm Challenge'' in which players could copy and optimize freely. We do not have the statistical basis to make general statements about these results and we therefore do not want to place too high emphasis on the obtained total atom numbers or the relative merit on the final convergence properties of players versus experts.

\subsection{Experimental details -- The Alice Challenge}

As described in the main text, the players control the loading sequence as well as the evaporation process through the game interface. In order to account for the high initial and low final parameter values of laser beam intensities and magnetic field gradients, the displayed ramps are represented on a logarithmic and normalized scale. In the Alice Swarm Challenge submitted solutions are placed in a waiting queue. Depending on the length of the queue, an estimated process time is displayed. This allows players to join, submit a single or multiple solutions and come back at a later time to review the achieved score. All results are placed on a high score list and  the players have the possibility to investigate and copy corresponding solutions completely or in parts. This facilitates reproducing working solutions and encourages the players to improve them further.

\subsection{Analysis of solutions  -- The Alice Swarm Challenge}
\label{SIF}
Due to the problem representation, player solutions feature, in general, a much smoother transition from loading to evaporation than the RedCRAB solutions or the parameter scans with a fixed loading sequence. Knowing the levitation gradient of \Rb in $\ket{F=2,m_F=2}$, one can estimate when the loading of a given trap configuration is finished. In these terms, the loading in the case of the best performing player solutions happens within about \SI{1}{\second}. This is about twice as fast compared to the standard loading sequence described above. 
Afterwards, the magnetic field gradient is lowered only very slowly and remains nearly constant just below the levitation gradient. This value is about \SI{70}{\percent} higher compared to the HT or best performing RedCRAB solution. Only in the last second of the sequence is it relaxed to a value similar as in the HT. 
The intensities of the dipole beams are lowered after \SI{1}{\second} which is another indicator for the transition from loading to evaporation. Compared to the HT, extremely low intensities are reached at the end of the sequence and it seems that current experimental conditions, such as beam overlap and alignment, are optimally employed.

As explained in the main text, the Alice Swarm Challenge was restarted multiple times, thereby varying $T_\mathrm{ramp}$. Investigating the quality of solutions as a function of pre-set $T_\mathrm{ramp}$, we found that for ramps below \SI{3}{\second}, $N_\mathrm{BEC}$ decreases drastically and no solutions were found for $T_\mathrm{ramp}<\SI{1.75}{\second}$. The shape of the ramps for long and short durations are considerably different. For instance, the initial beam intensities are much higher for the short-duration player solution. This leads to a stronger confining trap accommodating the shorter ramp duration.

%-----------------------------------------------------
  % Figure sweep the ramp duration for RedCRAB and Alice Challenge solutions
\begin{figure}[htb]
	    \centering
	    \includegraphics[width=0.85\linewidth]{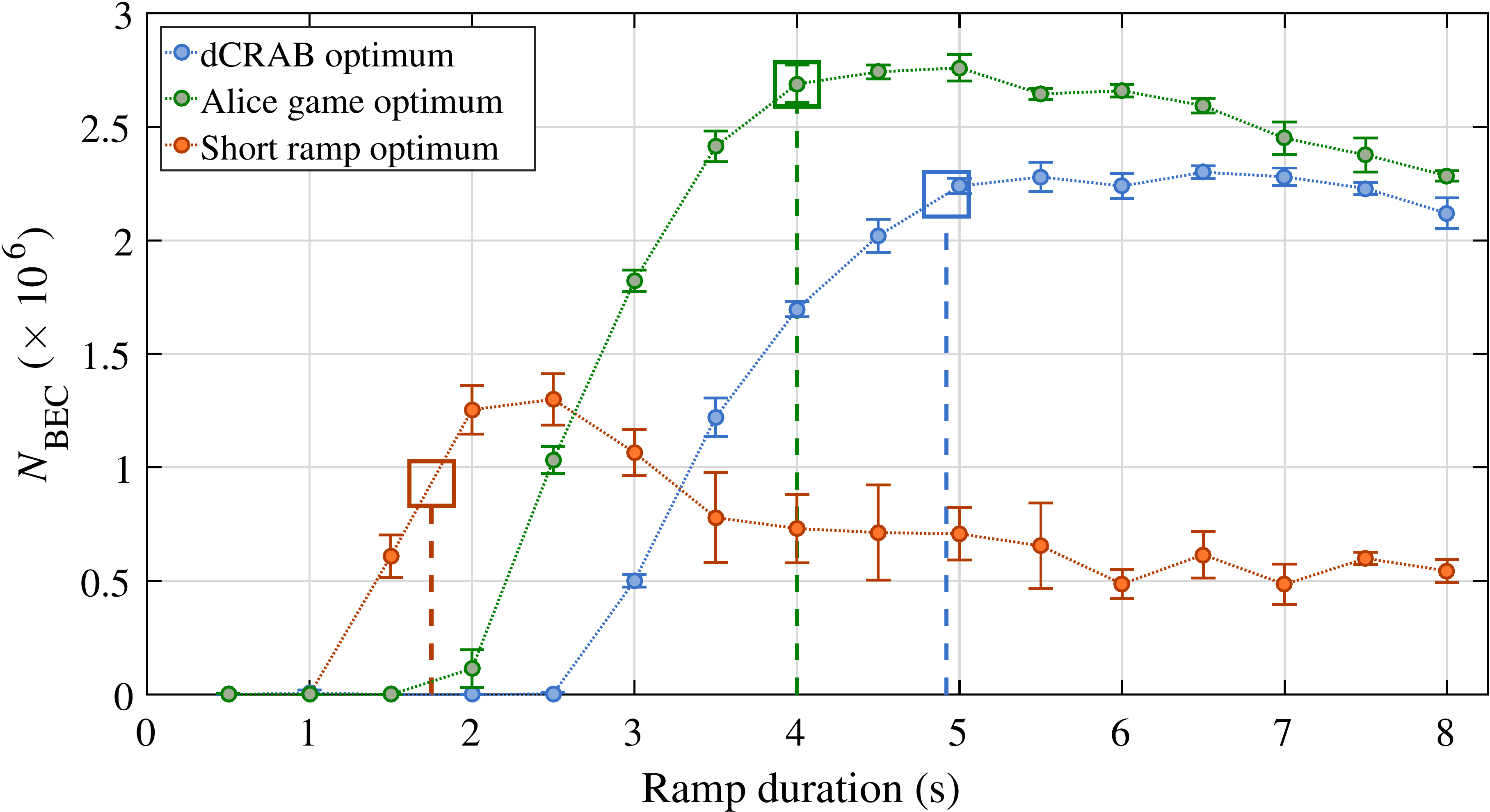}
		\caption{Sweep of the ramp duration, $T_\mathrm{ramp}$, for different optimum solutions. The ramp shapes yielded through the RedCRAB optimization and from two Alice Swarm Challenge sessions were scanned as a function of total ramp duration and $N_\mathrm{BEC}$ measured. As Alice Swarm Challenge solutions, the ramps resulting in the largest $N_\mathrm{BEC}$ and the solution for the shortest set ramp duration were chosen. The data points are obtained by averaging over five repetitions, where the error bars represent the standard deviation. The big squared marks denote $T_\mathrm{ramp}$ during optimization. Note, that the RedCRAB algorithm's control was restricted compared to the one of the players. For details, see text.}
		\label{fig:alice_sweep_solutions}
\end{figure}
%-----------------------------------------------------

As a concrete example of the difference between obtained solutions, we compare the duration-robustness of the RedCRAB solution and player solutions obtained at short and long durations ($T_\mathrm{ramp}=\SI{1.75}{\second}$ and \SI{4}{\second}, respectively) by stretching and compressing the solutions in time. Figure~\ref{fig:alice_sweep_solutions} demonstrates that solutions obtained at different durations exhibit different behavior. The features that are optimal for short ramps seem to be suboptimal for longer durations. Additionally, the scan reveals that the long-duration player solution is more robust than the RedCRAB solution against reductions in duration. We stress that since it was not a part of the optimization criteria, this behavior is not a reflection of the relative merit of the two methods of optimization. Rather, it highlights the strength of identifying a diversity of solution strategies. This not only gives information about the optimization landscape, but is also potentially an experimental advantage if retrospectively one needs to consider a more complex fitness function such as constraints on ramp duration or laser intensity.

\subsection{Individual and collective problem-solving -- The Alice Team Challenge}
The fact that natural science challenges \cite{Cooper2010SI,Sorensen2016SI,Lintott2008SI,Fischer2012,Lee2014cSI,Kim2014SI} can be solved efficiently by the general public is of interest to cognitive and social scientists as a source of insight into the general process of human problem-solving. However, data from these projects suffer from the fact that they were not gathered with particular social and cognitive science research questions in mind. Here, we set out a new kind of social science approach, where we investigate \textit{how} a collective of citizens is able to balance local vs. global search in a real-world setting. Such insight is important for future designs of large-scale citizen science investigations of natural science problems and advance general understanding of the process of individual and collective problem-solving.

On an individual level, experiments in the lab have shown that individuals adapt their search based on performance feedback \cite{billinger2013searchSI} and their search strategies are thus not merely local, nor global \cite{billinger2013searchSI,Vuculescu2017SI}. Specifically, if performance is improving, humans tend to make smaller changes (i.e. local search), while if performance is worsening humans tend to make larger changes (i.e. search with a global component). Studies in the field of cultural evolution have established that some of the most efficient social learning strategies are to copy the best or most frequent solutions and copy when the situation is uncertain \cite{morgan2012evolutionarySI,muthukrishna2016andSI,Mason2012SI}. Due to these social learning strategies, collective search is acknowledged to be effective at boosting the efficiency of a search process \cite{rendell2010copy}.

However, potentially constrained by the low dimensionality of the tasks to be solved, with few exceptions \cite{derex2015social} prior studies on human problem solving have primarily focused on simplified situations where individuals have the option to either copy another solution or not \cite{laland2004social}. As argued in \cite{Smaldino2012}, research should not only study option selection, but option generation where participants are not constrained by relatively few options, but are allowed to integrate and transform individual and social search information. This enables analysis of \textit{how much} individuals are influenced, rather than merely \textit{if} they are influenced.

Another key characteristic of previous studies is that they have primarily relied on a particular type of students at particular universities in a lab setting \cite{henrich2010}. Furthermore, this work has relied on these participants solving artificially designed tasks. Researchers have usually designed the problem task to be solved, and therefore also the nature of the landscape to search and what constitutes a good solution (see e.g. \cite{derex2015social, Mason2012SI, billinger2013searchSI, Vuculescu2017SI, acerbi2016socialSI, morgan2012evolutionarySI, weizsacker2010we}). Our experimental framework points towards a possible solution to these challenges, since we investigate how citizen scientists, engaged in a real world, high-dimensional and mathematically well-defined problem, adapt their search strategies to performance feedback and inspiration from solutions of other players. Finally, while individuals' ability to do adaptive search may make them uniquely suitable for navigating rugged fitness landscapes as evidenced in previous work \cite{Sorensen2016SI,Kearns2006,Judd2010}, we do not know how this adaptive search mechanism plays out in a \emph{collective} search environment. 

\subsection{Experimental treatments -- The Alice Team Challenge}
\begin{figure*}[ht]
    \centering
	\includegraphics[width=0.8\linewidth]{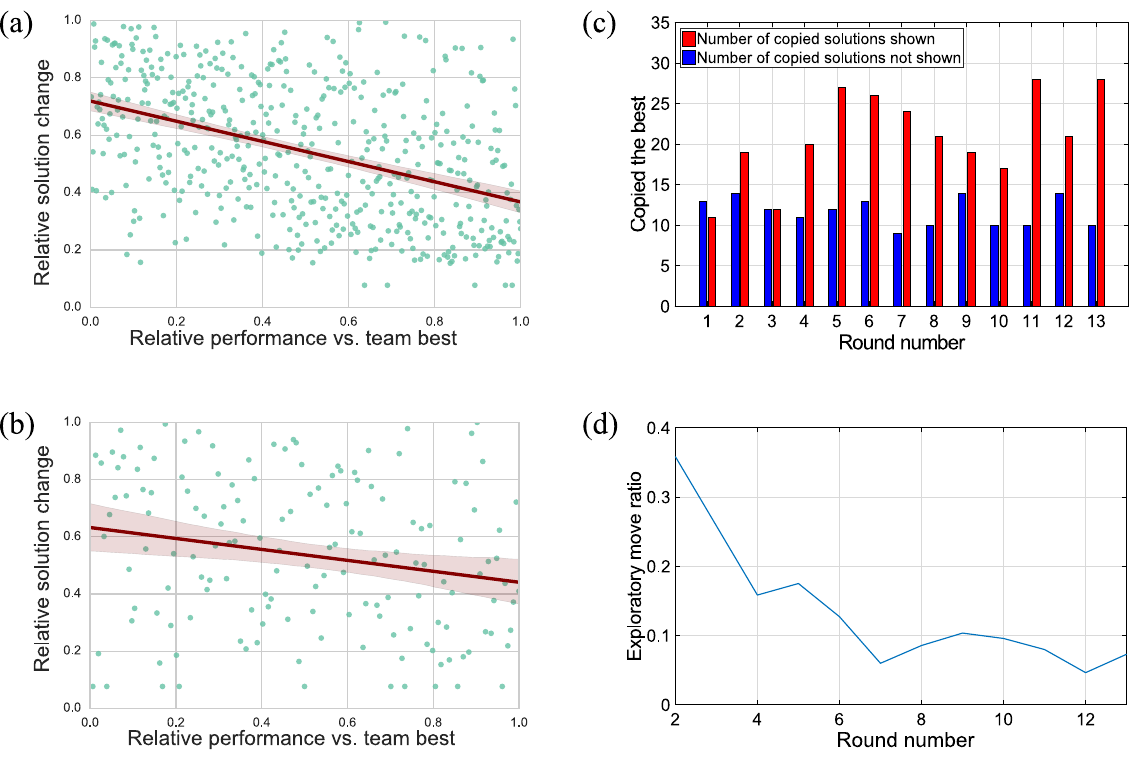}
	\caption{(a) and (b) How much players edit their own solution compared with relative performance in the previous round. (a) only includes submissions that did \textit{not} involve any kind of copying, whereas (b) shows the data that involved copying another solution. In both cases, the solutions of players that performed well relative to the team are changed less than players who did not perform well. Both the distance and score measures are ranked within each round with the current team-best score as a reference point. A \SI{95}{\percent} confidence interval is shown. The distance measure is based on distance from the players own previous solution (a) and on distance from copied solution (b).  (c)~Participants in the condition where the meta-information of seeing the number of copied solutions in their team in the previous round was available (red bars), used the ``copy the best'' social learning strategy more than participants in the condition where this information was not visible (blue bars). A two-sample t-test showed the two to be statistically different (p<0.0001). (d)~Variation of exploration in time. The measure for exploration was derived relative to the entire solution space covered by the players in the ATC. By computing distances from any two solutions submitted in the challenge we have obtained an average distance step of $\sim 6.13$.}
	\label{fig:alice_team_compilation}
\end{figure*}

We randomly allocated participants into two different experimental conditions, in order to further investigate the collective search process. In the treatment condition, information regarding how often a certain solution was copied from the previous round by team-members was available to participants, while in the control such information was not available. We conjectured that by presenting participants with this information, we could test the occurrence of explicit metacognitive social learning strategies \cite{heyes2016knows}. More specifically, we hypothesized that because participants could see how many of their team’s solutions had been generated by social learning (as opposed to individual search not involving copying) they could compensate, among others, for \textit{under-reliance} on social learning and copy \textit{more} in a given round. In this way we take the first step to study if a collective of human searchers can function as an adaptive global search algorithm, gradually changing their recombination intensity according to their performance as well as the meta-information received.
Figure~\ref{fig:alice_team_compilation}c demonstrate the results. Participants in the condition that were exposed to meta-information about how often a solution had been copied in their team, engaged in more social learning than participants in the condition where this information was not available. This shows that it is possible, via a simple manipulation, to nudge the human players into relying more on social learning strategies, specifically ``copy the best'' (see also table~\ref{t:strategiedistribution}). Considering that previous work \cite{weizsacker2010we, mcelreath2005applying} argues that human solvers rely \emph{too little} on social learning, and thus that an increase in relying on social learning strategies is desirable, this is a promising result. In the following we provide an analysis of the aggregated search behavior, in order to establish if and how a collective of humans are able to solve such a high-dimensional problem.

\subsection{Analysis of individual and collective search -- The Alice Team Challenge}
\label{SII}
\begin{figure}[ht]
    		\centering
	 		\includegraphics[width=0.8\linewidth]{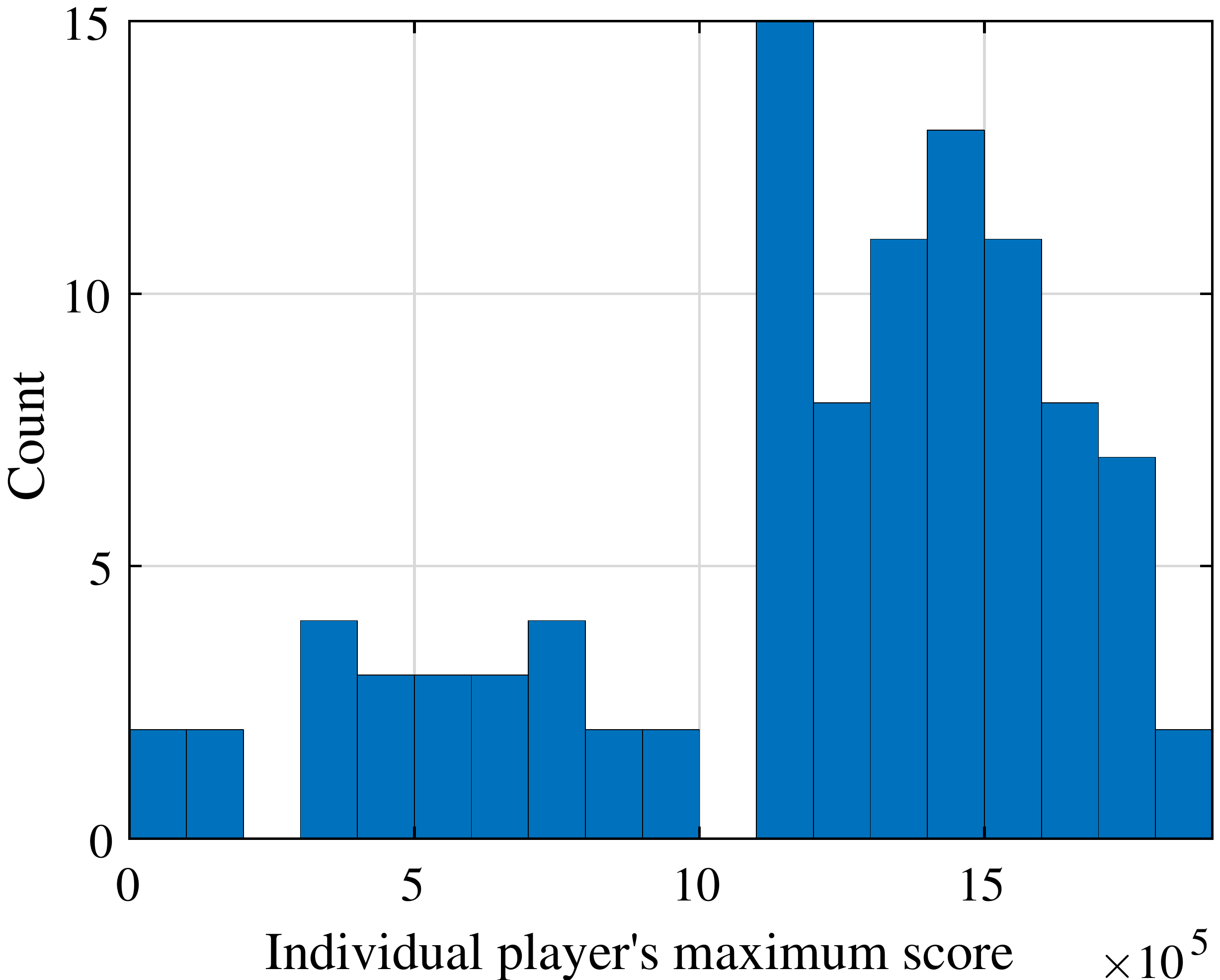}
	 		\caption{Histogram of individual highest achieved scores in the ATC. Overall, despite a very restricted number of tries, human players achieve relatively good scores. This histogram illustrates that only a handful of players have very low scores, while most players achieve scores above the 1 million threshold.}
	 		\label{SIfig7}
\end{figure}

Almost all teams and individuals managed to submit relatively good solutions above the 1 million threshold (see figure~\ref{fig:alice_redcrab_results}c in main text and figure~\ref{SIfig7}). Similar to Ref.~\cite{derex2015social}, we find that \SI{53}{\percent} of all moves were individual, \SI{41}{\percent} involved some form of social learning and the remaining \SI{5}{\percent} were random, \ie randomly recombining existing solutions (see table~\ref{t:strategiedistribution}). As outlined in the main text, we test and are able to show that players adapt their search based on the performance feedback they receive (see figure~\ref{fig:alice_redcrab_results}e where feedback and distance measures are ranked across all players and rounds). To further support this analysis, we normalize all scores in the following modelling efforts. We analyze the data using generalized linear mixed models with a Gaussian (or Binary with a logit function, where appropriate) error structure, and we control for individual variance by allowing for a random subject effect. This approach allows us to estimate a generalized model of adaptive search where individual heterogeneity and the repeated nature of the measures are taken into account. Models are constructed by forward inclusion and reported effects are within a \SI{95}{\percent} confidence interval.

We model the distance from the players' latest solutions as a function of the feedback players received in the form of a score. In this way we track how players responded to information about the underlying landscape. We find players are more likely to make minor changes to their solutions when they achieve a performance comparable to the team-best versus more significant changes when their performance is low comparative to team best feedback (CI: (-1.61: -1.21), p<0.0001). This supports former findings on individual adaptive search \cite{billinger2013searchSI, Vuculescu2017SI}. Additionally, players tend to become more conservative as the rounds progress (CI: (-0.12:-0.06), p<0.0001), see also figure~\ref{fig:alice_team_compilation}d).

Our collective setup also allows an investigation of \textit{if} and \textit{how} players engage in collective adaptive search, i.e. if they not only modify their own solutions, but also actively copy the solutions of other players (social learning). First, we establish that whether players engage in social learning or not depends on previous performance, where low performance leads to a higher likelihood of engaging in social learning (CI: (-0.92: -0.41), p<0.0001), see also \cite{muthukrishna2016andSI, derex2015social, acerbi2016socialSI, morgan2012evolutionarySI}. We also studied social learning in terms of which peer solutions players copy. This behavior follows a similar adaptive mechanism: players will tend to copy more dissimilar solutions, provided they had just experienced low performance (CI:(-1.35:-0.36), p<0.0001). Conversely, with low score differences, players are more likely to copy solutions similar to their own.

Furthermore, advancing previous studies, our setup allows us to investigate how players manipulate solutions \emph{after} engaging in social learning (\ie copying another solution), but before submitting their solution. As in pure individual search, players behave adaptively in this social situation: if they performed better than the previous team-best, their submitted solution will tend to stay closer to the copied solution (CI: (-2.33: -1.34), p<0.0001). Conversely, if their past score is below the same benchmark, their submitted solution will tend to drift further away. Interestingly, a comparison of the rank-based slopes in the graphs illustrating individual and social search shows that the adaptive effect performance induces in subsequent search, appears stronger for individual search than search that involved social learning (see figure~\ref{fig:alice_team_compilation}a and Fig.~\ref{fig:alice_team_compilation}b for details).

Overall, we created a novel, online gamified interface connecting a real-time physics experiment to citizen scientists. The setup provided a unique opportunity to both measure how much a collective of players change their solutions but also have an external measurement of the quality of the solution in the solution space. This enables one going beyond merely claiming human superiority \cite{Cooper2010SI, Sorensen2016SI} and study \textit{how} human problem solvers are efficient at balancing the trade-off between global and local search.

We show how individuals search adaptively depending on their own former performance, thus supporting lab-based studies based on artificial, low-dimensional problems \cite{billinger2013searchSI, Vuculescu2017SI}, while simultaneously expanding this adaptive mechanism to the realm of social learning. Even though the nature of the adaptive search mechanism is the same for both individual and social search, we find exploratory evidence for social search inducing less conservativeness for high performers. Finally, our innovative experimental game setup allows a genetic algorithm inspired opportunity to recombine and manipulate existing solutions, going beyond a simple imitation option in each round that simpler setups were constrained by \cite{rendell2010copy, weizsacker2010we}. When searching for an optimal solution in a complex, high-dimensional problem space, our exploratory investigation shows that humans don’t indiscriminately copy other solutions. They often only copy part of the solution and then further transform the copied solution in an adaptive manner. The fact that these individual and social adaptive search mechanisms systematically depend on the individual searchers' relative performance creates a diverse mixture of search within a collective, shaping the collective balance of local vs. global search and when the collective stops searching.

\subsection{Variables used in the analysis of individual and social search -- The Alice Team Challenge}
Feedback is defined as the ratio between the individual’s previous score and the best team score recorded so far. For individual adaptive search, a similar analysis was conducted using different benchmarks for feedback (i.e. either individual best or second to last submission). These models yielded qualitatively similar results. In the social adaptive learning situation only the social team-best score was considered to be a relevant benchmark. This operationalization follows previous studies that show individuals benchmark their performance against the best performance so far \cite{billinger2013searchSI, Vuculescu2017SI}. 
The search distance is given by the Euclidean distance between consecutively submitted solutions, in the case of individual adaptive search. For social adaptive search, the similarity variable refers to the distance between the copied solution and the solution submitted in the previous time step by the player. The shorter the distance between two solutions, the more similar they are. In figure~\ref{fig:alice_redcrab_results}e of the main text, figure~\ref{fig:alice_team_compilation}a and figure~\ref{fig:alice_team_compilation}b feedback and distance measures are rank-based across all players and rounds, while scores are normalized in all modelling efforts. For the reported analysis, the feedback scores are normalized by dividing the individual’s current score with the team best so far in the game, leading to corresponding numbers between 0 and 1. Results are robust to varying modelling assumptions, such as standardizing the data, controlling for individual heterogeneity within round or taking round (time) as a fixed effect.

\subsection{Characteristics of team-participants -- The Alice Team Challenge}
At the end of 13 rounds, participants were redirected and asked to fill in a brief survey. This allowed us to collect a number of demographic variables about the players. We had a response rate of \SI{80}{\percent} (89 respondents). The majority of our players were male (\SI{69}{\percent}), with an average age of 30.1 ($\mathrm{st.d}=10.12$). With respect to education, \SI{66}{\percent} of the participants had obtained a higher education degree, and a little over half (\SI{53}{\percent}) had physics as a subject in their education, after high school. Respondents were from 17 different countries, the majority being from Europe or North America.

\begin{table}%[tbhp]
\centering
\caption{Search strategies in the ATC}
\label{t:strategiedistribution}
\begin{tabular}{lrlr}
\toprule
\multicolumn{2}{c}{Overall search strategies} & \multicolumn{2}{c}{Social learning strategies}\\
\colrule
Individual search & \SI{53.4}{\percent} & Copy the best & \SI{57.6}{\percent} \\
Social learning & \SI{41.4}{\percent} & Other copying behavior & \SI{42.4}{\percent} \\
Shuffle & \SI{5.1}{\percent}  &  & \\
\botrule
\end{tabular}
%\par 
\justify\small
The table shows how often players engaged in various types of search moves. Individual search refers to moves that did not involve any form of copying. Social learning strategies refer to moves that involved copying others. Copy the best refers to copying the entire solution (all three lines) that at the time was the best. Other copying behavior refers to copying any other solution or only partially copying the best (\eg one or two of the three available lines). If a player ``shuffled'' he received a random combination of existing solutions, i.e. each line could be from different solutions.
\end{table}

%\bibliography{library_strategies_in_physics,pnas-zotero}

\begin{thebibliography}{57}
\bibitem [{foo(2017)}]{footnote_tsne}%
  \BibitemOpen
  \bibinfo {note} {Note, the
  displayed data points stem from an older set of measurements, where relative
  $N_\mathrm{BEC}$ between the trap configurations were not represented
  correctly. Therefore, the labelling of the color scale was
  omitted.}\BibitemShut {Stop}%
\end{thebibliography}
\begin{thebibliography}{45}
\bibitem [{foo(2017)}]{footnote_error}%
  \BibitemOpen
  \bibinfo {note}{The error bounds
  represent the standard deviation and are based on benchmarks taken randomly
  during one month's period. However, all our experiments presented in the
  following were conducted over the span of six months and long term drifts of
  the experimental apparatus were present. RedCRAB optimization and the
  parameter scans spanned over five months, the Alice challenge and
  experimental investigation of specific solutions over one month. Individual
  trap configurations were therefore subjected to an overall drift of up to
  \SI{14}{\percent}. Despite this, the relative yield of different strategies
  has been investigated with low noise within a short time frame, and over
  longer time scales by relating to benchmark sequences.}\BibitemShut {Stop}%
\end{thebibliography}

\end{document}